\documentclass{aa}  

\usepackage{graphicx}
\usepackage{txfonts}
\usepackage{xcolor}
\usepackage{placeins}
\usepackage{subcaption}
\usepackage{longtable}
\usepackage{array}     
\usepackage{amsmath}   
\usepackage{siunitx}   
\usepackage{afterpage}
\usepackage{gensymb}
\usepackage{hyperref}
\usepackage{threeparttable}
\begin{document}

   \title{Halo stars harbour few wide ultracool companions \thanks{\email{junyan.zhang@uwo.ca}}}

   \author{
  J.-Y.\ Zhang 
  \inst{1,2,3,4}        
     \and  N. Lodieu \inst{1,2}
     \and  E.\ L.\ Mart\'in \inst{1,2}
        }

   \institute{Instituto de Astrof\'isica de Canarias (IAC), Calle V\'ia L\'actea s/n, 38200 La Laguna, Tenerife, Spain
       \and
       Departamento de Astrof\'isica, Universidad de La Laguna (ULL), Avenida Astrofísico Francisco Sánchez s/n, 38206 La Laguna, Tenerife, Spain
        \and
       Department of Physics and Astronomy, The University of Western Ontario, 1151 Richmond St, London, ON N6A 3K7, Canada
       \and
    Institute for Earth and Space Exploration, The University of Western Ontario, 1151 Richmond St, London, ON N6A 3K7, Canada
       }

   \date{Received 30 January 2026 / Accepted 28 May 2026}

 
  \abstract
   { 
    }
   {We aimed to provide a wide ultracool companion frequency for metal-poor halo stars, for the first time.}
   {We selected nearby halo stars with high proper motions and metallicities determined spectroscopically. We collected first-epoch deep $J$-band imaging around these stars. We visually compared these images with optical images from public surveys, and those that have nearby faint sources beyond optical survey limit were followed up by deep $J$-band imaging with baselines of two to four years. We searched for comoving sources between the two epochs.}
   {Our observation reached an average depth of $J_{\mathrm{lim}}=22.8$ and 23.0\,mag (Vega) in the first and the second epoch, respectively, enabling us to detect extreme subdwarfs earlier than type esdT0 at a distance up to 250\,pc. With this depth, we did not find any bona fide wide ultracool comoving companion to a nearby metal-poor halo star within a typical projected separation range of a few hundred to a few thousand au from the star. We put an upper limit of the frequency of wide ultracool companions to metal-poor halo stars of 4.0\% at a 90\% confidence level. With four stellar comoving companions found and confirmed by \textit{Gaia}, we derived the wide stellar companion frequency of $6.1^{+7.2}_{-4.0}\%$ at a 90\% confidence level.}
   {We conclude that wide ultracool companions are rare around metal-poor halo stars with a frequency marginally lower than that of wide stellar companions. So far, we claim that no metallicity dependence is found for the wide ultracool companion frequency around stars. Formation and retention processes in binary systems are likely to operate less efficiently for ultracool secondaries.}

               
\titlerunning{}
\maketitle 
%

\section{Introduction}

Ultracool dwarfs (UCDs) are objects with spectral types later than M7.0\,V and effective temperatures $T_{\mathrm{eff}} \lesssim2700$\,K \citep{kirkpatrick1997ucd,kirkpatrick2005LT}. Owing to their inefficient or absent sustained hydrogen fusion, halo UCDs serve as valuable fossil tracers of the early Milky Way. Consisting of pristine materials, halo UCDs are metal-poor, thus serving as natural laboratories for testing low-metallicity cold atmosphere models. According to their spectral features, they are classified into three metallicity subclasses: subdwarf (sd), extreme subdwarf (esd), and ultra subdwarf (usd) \citep{lepine2007sdM,zhang2017six_sdL_classification,burgasser2025esdT}. Late-type M subdwarfs started being detected in the first generation of digitized surveys almost three decades ago \citep{gizis1997pm_subdwarf,scholz2004sdM9,lodieu2005sdM}, followed shortly thereafter by the discoveries of L subdwarfs \citep{burgasser2003esdL7_2M0532,lepine2003first_early_sdL}.

Thanks to the well-constrained properties of the stellar primary and the wide separation that enables the secondary to be resolved, halo ultracool subdwarfs as wide companions to stars or evolved stars are considered valuable benchmarks. Several such systems have been identified in recent years; for example, HD\,114762B as an late-M subdwarf wide companion to an F subdwarf HD\,114762A \citep{bowler2009hd114762B}, Wolf\,1130C as a late-T subdwarf wide companion to an M subdwarf-white dwarf binary system Wolf\,1130AB \citep{mace2013Wolf1130C,mace2018Wolf1130,burgasser2025phosphine}, GJ\,660.1B as an late-M subdwarf wide companion to an early-M subdwarf GJ\,660.1A \citep{aganze2016GJ660.1AB}, 
Gaia J045245.87$-$360843.8 
as an early-L subdwarf wide companion to an early-M subdwarf Gaia\,J045238.82$-$361001.3 \citep{zhangzenghua2019MLbinary}, and VVV\,J125641.09$-$620203.8 
as an early-L subdwarf wide companion to a white dwarf VVV\,J125644.42$-$620208.1  
with a progenitor mass of $1.9\pm0.4$\,M$_\odot$ \citep{zhangzenghua2019sdL3,zhangzenghua2024sdL+WD}.

Although a small number of halo wide ultracool companions to stars have been confirmed, the wide ultracool companion frequency in the halo stellar population remains poorly constrained, and how metallicity would affect the formation of these systems is unclear. Halo stars account for a small portion of the whole stellar population, hence it is necessary to cover a large volume to have a statistically significant sample. However, the intrinsic faintness of UCDs, especially that of ultracool subdwarfs, prevents us from detecting them at large distances.
In comparison, several studies have investigated wide stellar companions to metal-poor stars and found frequencies ranging from a few percent up to $\sim$10--20\% \citep[][]{zapatero2004metalpoor_wide_binary,jao2009cool_subdwarf_companion_speckle,zhang2013SDSS_sdM_binary,ziegler2015cool_subdwarf_companion_highres,hwang2021binary_metallicity,lodieu2025widebinary}. Others have reported wide ultracool companion frequency of solar-metallicty stars and young stars of a few percent \citep[][]{metchev2009companion_mass_func,chinchilla2019companion_young,dalponte2020ultracool_binary}.

To fill this gap and provide valuable constraints on the formation and evolution of such systems, we carried out a dedicated survey to identify comoving wide ultracool companions to metal-poor stellar primaries. It involved two-epoch deep imaging around a large sample of selected metal-poor halo stars. We obtained the images in the near-infrared (NIR) wavelengths where UCDs are relatively bright. 

This work is structured as follows: Section~\ref{obs} describes the sample selection, observation details, and data reduction. The main results are presented in Section~\ref{results}. In Section~\ref{discussion}, we compare our results with those from the literature and discuss their implications. Section~\ref{conclusion} summarises the work.

%
%
\section{Observations}
\label{obs}
%

%
%
\subsection{Sample selection}
\label{sample_selection}

We selected 66 bright halo stars observable from the Canary Islands, with spectroscopically determined metallicity [Fe/H] $<-1.5$\,dex, from the 1447 stars in the catalogue of \citet{carney1994haloFGK}. This catalogue was based on historical proper-motion surveys, implying that the selected stars exhibit sufficiently large proper motions to be detectable with earlier observational techniques. As a result, the sample consists of nearby, relatively bright, and highly reliable halo members. In addition, the use of high-resolution spectroscopy provides robust metallicity measurements.

To ensure that potential ultracool companions would be detectable, all of the sample are within 250\,pc, with seven are within 100\,pc.
These stars have effective temperatures from 4600 to 6300\,K.
The total proper motions of these targets range from 193 to 2204\,mas\,yr$^{-1}$, allowing measurable motions from the ground over the few-year baseline between the first and second epochs. The information about these 66 targets is listed in Table~\ref{targets}.

\subsection{First epoch observation}
\label{obs_1st_epoch}

For the first epoch, the observations were conducted under service programmes GTC53-20B and GTC65-21A (PI: N.\ Lodieu) during 2020 and 2021. Three targets were observed twice in this first epoch, which have two records in the first-epoch modified Julian date (MJD) in Table~\ref{targets}. We kept the images of higher quality between the two epochs.

We used the Espectrógrafo Multiobjeto Infrarrojo \citep[EMIR;][]{garzon2022emir} on the 10.4-m Gran Telescopio Canarias (GTC) located at the Spanish island of La Palma. EMIR is a common-user, wide-field, near-infrared camera installed on one of the GTC's Nasmyth foci. EMIR was equipped with a Teledyne HAWAII-2 HgCdTe near-infrared optimised chip of a size of 2048$\times$2048 pixels. The field of view (FoV) of EMIR is 6\farcm67$\times$6\farcm67 and the pixel scale is 0\farcs193\,pix$^{-1}$. The smallest total proper motion of the 66 targets corresponds to a minimum displacement of 3 pixels on the EMIR detector over approximately a three-year baseline between the first and second epochs. This motion is sufficiently large to be detected and measured with high significance and good precision ($>10\sigma$), as the centroid position error of the point spread function (PSF) for the faintest source at S/N $=3$ can be determined to about a quarter pixel with EMIR. 

We used the $J$-band filter because the $J$ band is less affected by strong collision-induced absorption in the low-metallicity ultracool atmosphere and suffers lower sky-background emission compared to the $H$ and $K$ bands. We deployed a standard seven-point dithering pattern with an offset of 25\arcsec, and the on-source exposure time for each object was 60\,s $\times$ 7\,dithering $\times$ 3\,cycles $=$ 1260\,s. Despite that all the primaries are extremely bright for a 10-m class telescope and would be saturated, the single exposure time was kept to 60\,s to avoid stray light saturating the surroundings of the primaries. The actual seeing conditions ranged from 0\farcs6 to 1\farcs2. There were no constraint on the moon phase, but the moon was required to be at least 30\degree\,away from the target.  

Two targets, G\,103-50 and G\,27-8B were not observed with the configuration, denoted with * in the first epoch MJD $t_1$ column in Table~\ref{targets}. The former one had very short exposure of $5\times5$\,s = 25\,s and the latter one did not have dithering and thus it could not have the background well subtracted. The images were not deep, but we still reduced them to extract as much information as we could.

\subsection{Data reduction}
\label{obs_DR}

The 2D EMIR frames were preliminarily reduced by the EMIR default pipeline {\tt PyEmir}\footnote{\url{https://pyemir.readthedocs.io/en/stable/\#}}. The pipeline performed bad pixel masking, flatfielding, sky subtraction using the dithering pattern, and stacking. 

To avoid distorted stars affecting the astrometry at the central region where we were interested, the astrometric calibration was applied after cutting the edge of the frames. 
We then used the {\tt Astrometry.net} script \citep{dustin2010astrometry.net} to solve the astrometry for the final stacked and cropped images.

\subsection{Target selection for the second epoch observation}
We visually examined the images to see if there were any close-by faint sources. We did cutouts of sizes of 90\arcsec $\times$ 90\arcsec centred at the stars. This cutout size was chosen to exclude sources at the edge that were substantially affected by instrumental aberrations and distortions. The projected physical sizes of the cutout should be $90 d$\,au $\times$ $90 d$\,au where $d$ is the distance of the star in parsecs. Thus, the maximum projected separation explored for each target ranged from a few hundred to a few thousand au. 

We compared these first-epoch images with Pan-STARRS \citep{chanbers2016panstarrs} coloured images with their red, green, and blue (RGB) channels corresponding to the $y$, $z$, and $i$ bands, respectively. 
We used these three reddest bands of Pan-STARRS to maximise the sensitivity of the ultracool objects. The sources that have Pan-STARRS detection would have been recognised moved significantly if they are comoving companions, thanks to the large baseline (6$-$11 years) between the Pan-STARRS's epoch and our first epoch. 

We recovered four stellar companions in the Washington Double Star Catalog: G\,79-56 with its companion LSPM\,J0341+0923W, discovered by Skiff B.A.; BD+00$^\circ$2058A with BD+00$^\circ$2058B, discovered by Herschel J.F.W.; G\,214-1 with G\,214-1B, discovered by \citet{zapatero2004metalpoor_wide_binary}; and G\,27-8 with G\,27-8B discovered by \citet{luyten1979NLTTcatalog}. Their common proper motions were confirmed by \textit{Gaia} \citep{gaiaedr3}.  We did not find any other faint sources that have red Pan-STARRS counterparts that share the proper motion with the primary. Therefore, for the second epoch observation, we selected 28 targets whose deep images exhibit close-by faint sources that were not detected by Pan-STARRS, which could be an ultracool comoving companion to the star. 

\subsection{Second epoch observations}
\label{obs_2nd_epoch}

For the second epoch in 2024, in total 21 out of these 28 targets with one repetition (G\,241-41) got observed with GTC/EMIR+ --- an upgraded GTC/EMIR with a new HAWAII-2RG detector with the same size, pixel scale, and orientation ---  under programme GTC45-24A (PI: N. Lodieu). Two out of these targets (BD+42$^\circ$2667 and G\,241-4) were observed during the commissioning of GTC/EMIR+ earlier on 30 August 2023.  The observational setup and data reduction procedures were kept consistent with those of the first epoch, with the exception that stacking offsets in the second epoch were determined automatically by the pipeline, rather than manually using the \texttt{imexam} task. The actual seeing conditions ranged from 0\farcs6 to 1\farcs0.

\subsection{Comovement detection}
\label{obs_comotion}

Figure~\ref{dual} presents the deep GTC/EMIR $J$-band images covering FoVs of 90\arcsec $\times$ 90\arcsec for all the 21 targets across both epochs. The images are displayed in a logarithmic scale to enhance the visibility of faint features, along with the corresponding difference images. The difference frames are significantly affected by background contamination caused by diffused light from the bright central star, making them suboptimal for identifying faint comoving sources. To overcome this, we also visually inspected the two-epoch images by blinking them in different scales (linear, logarithmic, asinh) using  {\tt SAOImage DS9}.

%
%
\section{Results}
\label{results}
We recovered four comoving stellar companions (indicated in Table~\ref{targets}). 
Only one comoving ultracool candidate was identified, whose primary is BD+02$^\circ$3375. This system was selected for further investigation.

\subsection{BD+02$^{\texorpdfstring{\circ}{\textdegree}}$3375}
\label{results_BD02}

\begin{figure}[tbp]
\centering
\begin{subfigure}[b]{0.46\textwidth}
    \centering
    \includegraphics[width=\textwidth]{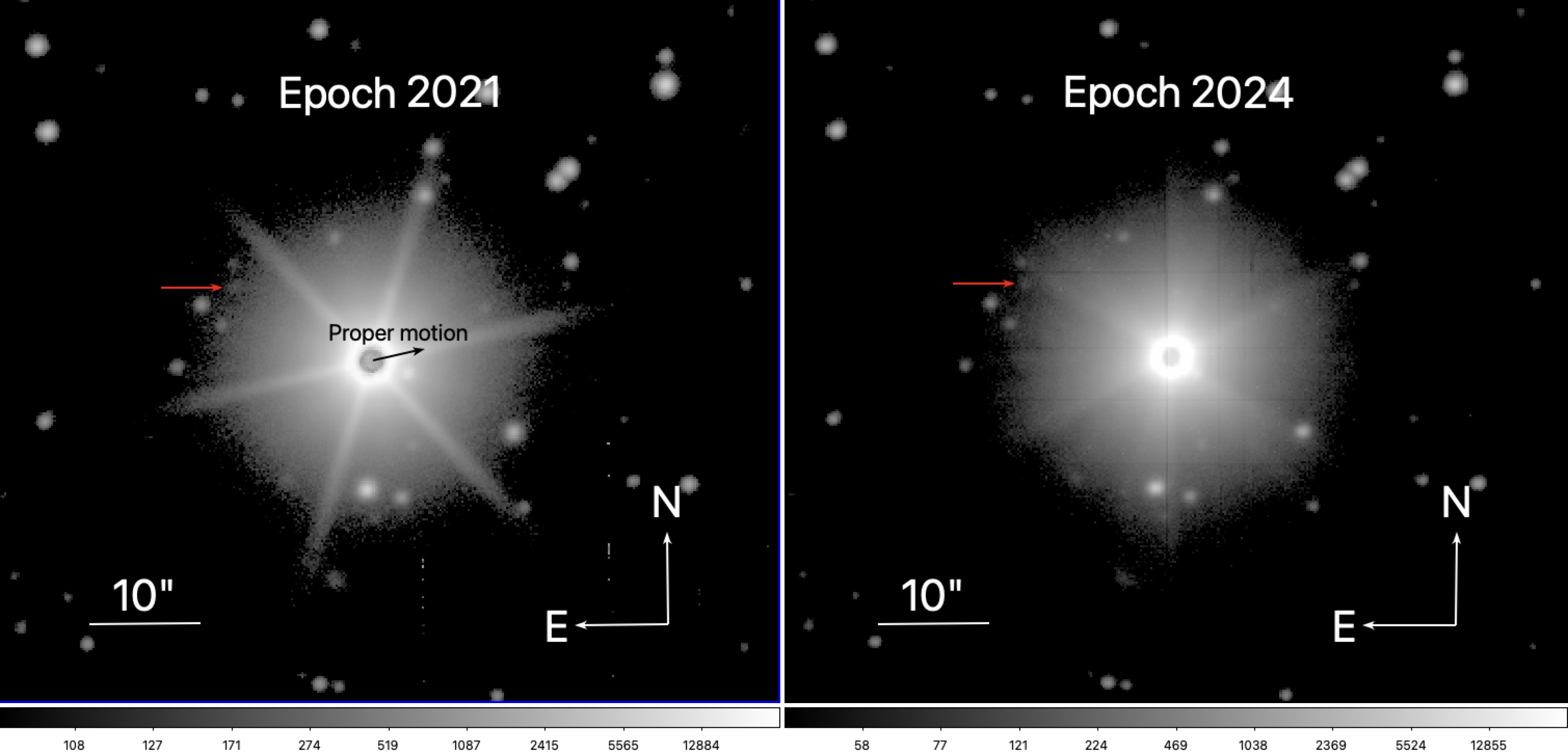}
\end{subfigure}
\begin{subfigure}[b]{0.47\textwidth}
    \centering
    \includegraphics[width=\textwidth]{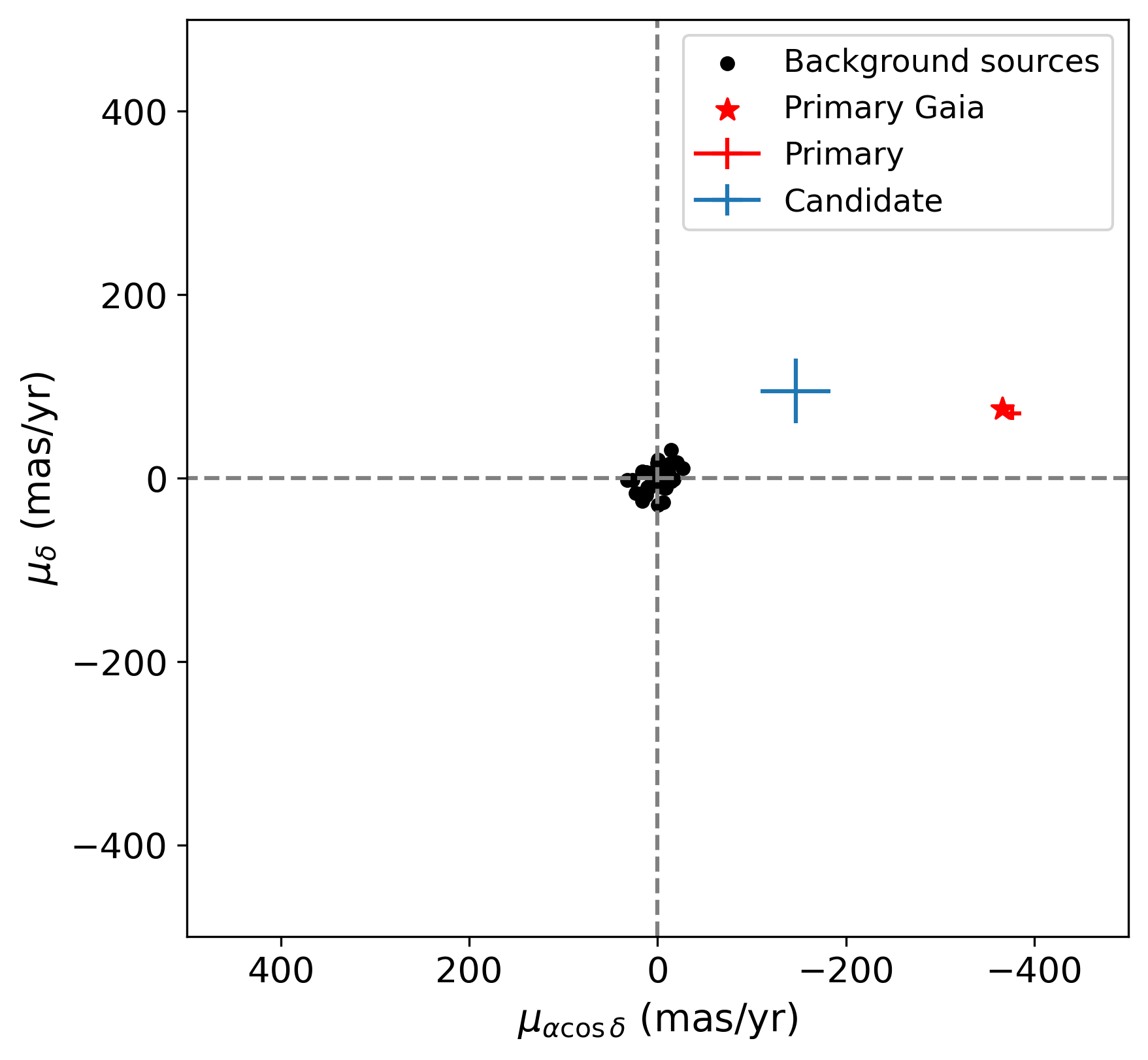}
    \end{subfigure}
\caption{Top: Two epochs of GTC/EMIR $J$-band images of BD+02$^\circ$3375 in logarithm count scale. The positions of the comoving companion candidate are pointed out by red arrows. The motion is not visible in the difference image of Figure\,\ref{dual}. Bottom: proper motion diagram of the sources. The proper motion of the primary (red cross) agrees well with \textit{Gaia} (red star) within the uncertainties. Background sources extracted by daofind and by hand have almost zero proper motion (black dots). The candidate (blue cross) has significant proper motion compared to background sources but is not comoving with the primary.}
\label{bd023375}
\end{figure}

We identified an extremely faint source at coordinates 17\textsuperscript{h}39\textsuperscript{m}45\fs87 +2\degree25\arcmin08\farcs26 (epoch MJD 60450.09), located north-east of BD+02$^\circ$3375, which appears to share the proper motion of the star (top panel of Fig.~\ref{bd023375}).
Instead of performing independent astrometric solutions for both epochs which would introduce astrometric uncertainties twice, we adopted a relative astrometry approach at the pixel level, using a single reference frame from one epoch. This method minimises the impact of instrumental aberrations and distortions, particularly given that the instrument configurations were nearly identical across both epochs.

We used Image Reduction and Analysis Facility \citep[IRAF; ][]{tody1986iraf} task \texttt{daofind} to extract bright field stars and then used task \texttt{xyxymatch} to create a list of matched coordinates (number of matched stars is normally a few hundred). We manually measured the faint source position using \texttt{imexam} and \texttt{imcentroid}. 
We rotated the image using \texttt{rotate} with the primary star as the rotation centre, and subtracted the rotated image from the original frame to test whether this approach could help leverage the effects of stray light and diffraction spikes around the primary.
Then we fit the shift, rotation, linear magnification, and aberrations of the centre part of both images using the task \texttt{geomap}. At the end a task \texttt{geoxytran} transformed all the coordinates of one epoch to the reference frame of the other epoch using the parameter derived by \texttt{geomap}.

Although it has a motion with respect to the background at a significance of 5\,$\sigma$, and is in a similar direction of BD+02$^\circ$3375, the proper motion ($\mu_{\alpha}\cos\delta=-146\pm37$\,mas\,yr$^{-1}$, $\mu_{\delta}=+95\pm36$\,mas\,yr$^{-1}$) is different from that of the star ($\mu_{\alpha}\cos\delta=-377\pm10$\,mas\,yr$^{-1}$, $\mu_{\delta}=+71\pm8$\,mas\,yr$^{-1}$; from EMIR astrometry) at a significant level of 6\,$\sigma$ (bottom panel of Fig.~ \ref{bd023375}). The criteria used by \citet{montes2018FGK+M} to distinguish physical (bound) from optical (unbound) systems are a ratio of the proper motion value difference to the primary proper motion $\mu$~ratio $<0.15$, and a proper motion position angle difference $\Delta PA<15\degree$. Our candidate pair has $\mu$~ratio $=0.60$ and $\Delta PA=22.4\degree$.  We hence discarded the comoving scenario.

\subsection{Completeness}
\label{results_completeness}
\subsubsection{Depth}
\label{results_depth}

\begin{figure*}[htbp]
    \centering
    \includegraphics[width=\linewidth]{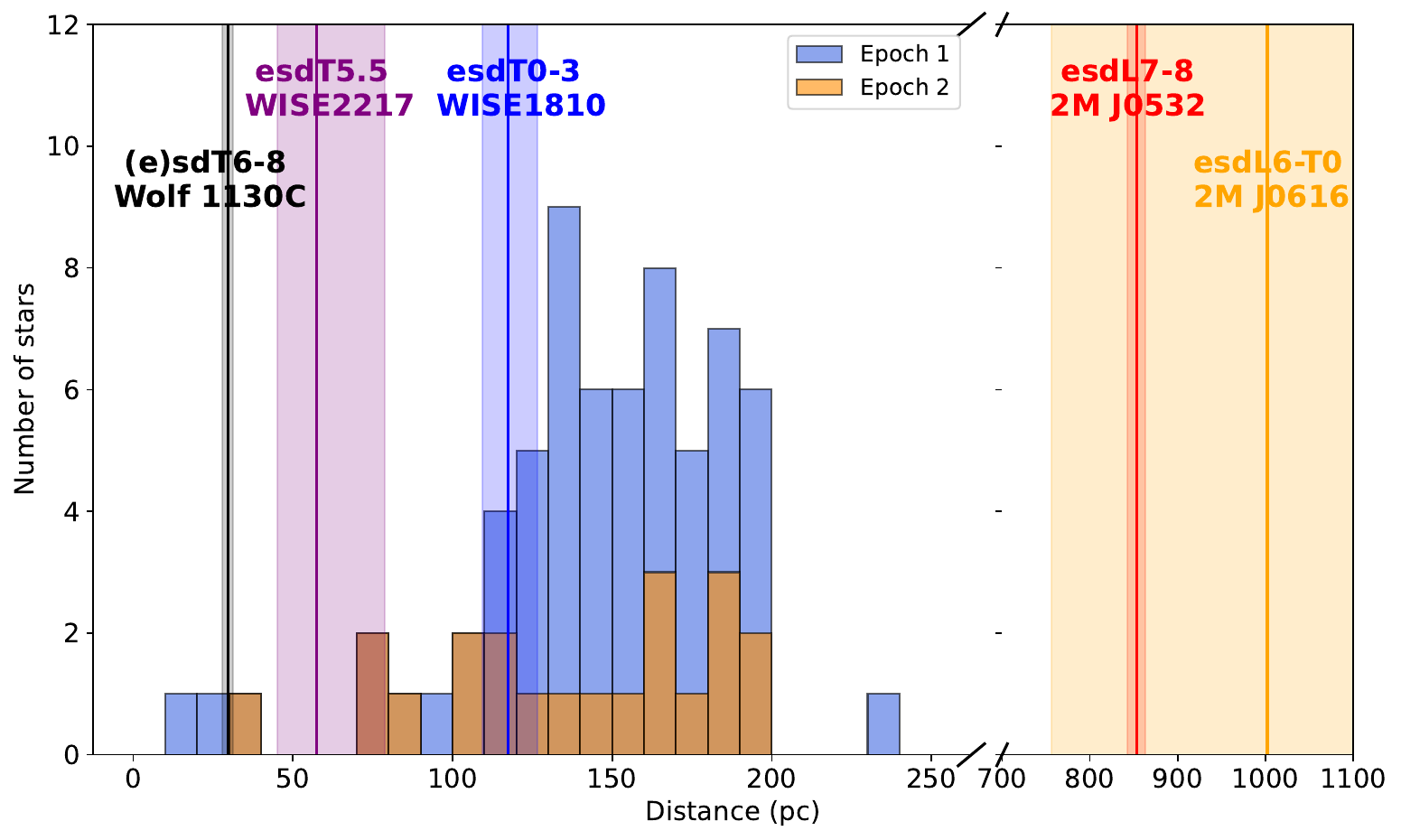}
    \caption{Target distance histogram compared with maximum detectable distances (coloured solid lines) for different spectral types of extreme subdwarfs with low metallicities at the first-epoch GTC/EMIR $J$-band depth of 22.8 mag. The coloured shades indicate the uncertainty of the maximum detectable distances, mainly result from the uncertainties of parallax measurements.}
    \label{distance}
\end{figure*}

We performed aperture photometry using {\tt Photutils} \citep{larry2024photutils} to determine the depth of the GTC/EMIR images. We measured the average full width half maximum (FWHM) of each image by fitting a 2D Gaussian profile for each source extracted by {\tt DAOStarFinder}, which is based on the algorithm by \citet{stetson1987daophot}. 
We used apertures with radii of 1.2 times of the FWHM of each image, and sky annuli with inner radii of 4\arcsec and outer radii of 6\arcsec. As photometric references, we selected all 2MASS sources with $J$-band magnitudes fainter than 14.0\,mag and located within 70\arcsec of the image centres.

For images containing at least four such reference stars, we determined the 3\,$\sigma$ limiting $J$-band magnitude by estimating the background fluctuation within a sky aperture of the same size. In the first epoch, our GTC/EMIR images reach a 3\,$\sigma$ $J$-band limiting magnitude of 22.8\,mag on average. For the second epoch it is about 23.0\,mag,  likely due to both the detector upgrade and improved average seeing conditions.

To estimate the latest spectral type of metal-poor UCDs detectable at the GTC/EMIR limiting magnitude of the first epoch $J_{\mathrm{lim}}=22.8$\,mag, we shifted known extreme subdwarfs with trigonometric parallaxes and similar metallicities to distances where their apparent magnitudes match $J_{\mathrm{lim}}$ (Fig.~\ref{distance}). The sample includes the benchmark extreme metal-poor T dwarf WISEA\,J181006.18$-$101000.5 (WISE1810$-$10), another benchmark late-type T subdwarf Wolf\,1130C, the mid-type esdT CWISE J221706.28$-$145437.6 (WISE2217$-$14), and two late-type esdLs 2MASS\,J05325346+8246465 (2M\,J0532+82) and 2MASS\,J06164006$-$6407194 (2M\,J0616$-$64). 

The benchmark WISE1810$-$10 is the closest esdT \citep[$d=8.9^{+0.7}_{-0.6}$\,pc; ][]{lodieu2022W1810}, and has the most precise metallicity measured among its kind \citep[$\mathrm{[Fe/H]}=-1.7\pm0.2$\,dex; ][]{zhang2025constraint}. Given WISE1810$-$10's $J$-band magnitude of 17.3\,mag, and a spectral type determined from esdT0 \citep{Schneider2020W0414_W1810} to esdT3 \citep{burgasser2025esdT}, we could have detected esdT3 objects akin to WISE1810$-$10 at a maximum distance of  $d\times10^{\frac{J_{\mathrm{lim}}-J}{2.5}\cdot\frac12}=112.0^{+8.8}_{-7.6}$\,pc with the aforementioned mean depth, following the inverse-square law. The same calculation was done for the rest; the photometric uncertainty and the parallax uncertainty were all taken into account. Wolf\,1130C has metallicity and trigonometric parallax from its primary --- an M subdwarf --- and has been classified from sdT8 \citep{mace2013Wolf1130C} to (e)sdT6 \citep{burgasser2025esdT}. WISE2217$-$14 has trigonometric parallax of $48\pm13$\,mas and has signs of extremely low metallicity \citep{zhang2025constraint,zhang2025wise2217}. It has a spectral type of esdT5.5 \citep{meisner2023coldoldBD}. 2M\,J0532+82 was classified from esdL7 \citep{zhang2013SDSS_sdM_binary} to esdL8 \citep{burgasser2025esdT} and has a precise distance of 24.56$^{+0.28}_{-0.27}$\,pc from \textit{Gaia}. 2M\,J0616$-$64 has a large uncertainty on the spectral type classification from esdL6 \citep{kirkpatrick2010pm_survey2MASS,zhang2017six_sdL_classification} to esdT0 \citep{burgasser2025esdT} and also has an uncertain distance of 50$^{+24}_{-12}$\,pc \citep{faherty2012bdkp_70ucd_parallax}.

Figure~\ref{distance} shows that our observations are sensitive to all potential extreme subdwarf companions with spectral types earlier than esdT0 within 250\,pc. For early-type esdTs, the detection completeness is estimated at approximately one-third of the whole sample. The sensitivity is insufficient to assess the presence of late-type esdTs or esdYs across the majority of the sample.

\subsubsection{Spatial coverage}
\label{results_coverage}

\begin{figure*}[htbp]
    \centering
    \includegraphics[width=0.98\linewidth]{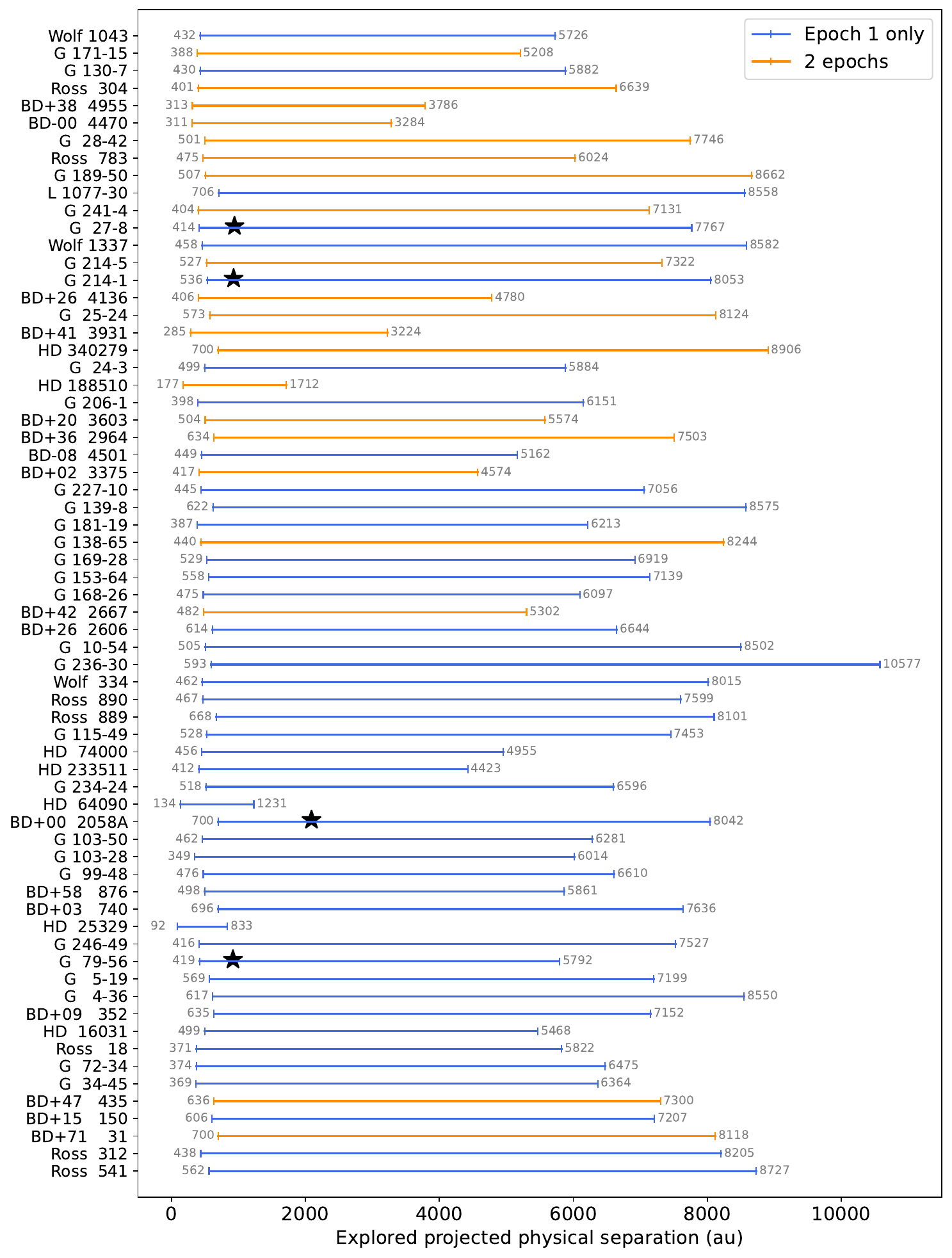}
    \caption{Explored projected physical separation ranges (au) for comoving companions of each target, with objects ordered by increasing right ascension from bottom to top. Objects with two epochs of GTC/EMIR observations are in orange and the rest are in blue. Black stars are the four identified known stellar companions.}
    \label{ranges}
\end{figure*}

All of the target stars are too bright to not saturate even in the 60\,s GTC/EMIR individual exposures. Considering that the stellar flux increases exponentially when the magnitude decreases, and the wing of the PSF approximately follows a 2D exponential profile, the saturation radius of the star is expected to decrease roughly as the square root of the magnitude. 
By checking on the image, we found that regions with an analog-to-digital unit (ADU) count $\gtrsim$ 30,000 lose the linearity. We empirically fitted the saturation radius in arcsec with the $J$-band magnitude of the star $R_{\mathrm{sat}}(J)\sim\left(13.0-J\right)^{\frac12}$.
This relation implies that stars fainter than 13.0\,mag in the $J$ band will not saturate in these exposures, consistent with the photometric analysis.

In practice, PSFs are more complex and are influenced by multiple factors, including atmospheric seeing and optical diffraction. 
To remain conservative in our analysis, we defined the innermost detectable separation for ultracool comoving companions as $2R_{\mathrm{sat}}$. While if $R_{\mathrm{sat}}$ is smaller than the worst seeing allowed in the observation, i.e., 1\farcs2, we instead fixed the innermost separation to $2\times1\farcs2=2\farcs4$. Given the distance to each star, this angular threshold translates to a physical projected separation in au. Combined with the outer boundary set by the cutout size, these limits define the range of separations that we probed for each target (Fig.~\ref{ranges}). Except for some very close-by and bright sources, the physical projected separation ranges from $\sim500$ up to $\sim8000$\,au from the central stars.

\subsection{Wide stellar companion frequency}
\label{results_frequency}
Four out of our 66 stars have comoving stellar companions confirmed by \textit{Gaia}, including G 27-8B that had failed to have deep image in the first epoch. Although \textit{Gaia} has a good completeness of M7 dwarfs at a distance of 100\,pc \citep{gaia2021nearbystar}, we calculated \textit{Gaia}'s limit for metal-poor subdwarfs. According to the absolute $G$ magnitude of M subdwarfs with metallicity [Fe/H] $< -1.0$\,dex derived from Table 4 of \citet{zhangshuo2021sdM_atm}, together with \textit{Gaia}'s limit of $G=20.7$\,mag \citep{gaia2016}, we inferred that \textit{Gaia} can detect an early-M-type extreme subdwarf at distance of 100--200\,pc. 

Mid- to late-M extreme subdwarfs at distances of $\sim200$\,pc, which are below \textit{Gaia}'s detection limit, can be readily detected by Pan-STARRS \citep{zhang2013SDSS_sdM_binary}. Nevertheless, none were found during the visual comparison between Pan-STARRS and our first-epoch observations.

Being complete until late-M extreme subdwarf companions, the wide stellar companion frequency of the halo stars would be $p_{\mathrm{star}}=k/n=4/66=6.1\%$. 
Assumed a binomial distribution to calculate the two-sided 90\% confidence interval (statistical significance $\alpha=0.1$) of $p_{\mathrm{star}}$, we have
\begin{equation}
    5\%=\frac{\alpha}{2}\le\sum_{i=0}^{k-1} \binom{n}{i} {p_{\mathrm{star}}}^{\,i} (1 - p_{\mathrm{star}})^{n - i} \le 1-\frac{\alpha}{2}=95\%,
\end{equation}
which yields $p_{\mathrm{star}}=6.1^{+7.2}_{-4.0}\%$.

\subsection{Upper limit of wide ultracool companion frequency}
\label{results_uplim}

Given that no faint UCD companions were confirmed comoving, we aimed to derive an upper limit on the true wide ultracool companion frequency $p$ at a 90\% confidence level, assuming the same binomial distribution as that for calculating the stellar companion frequency. 
Since two targets did not reach a depth for probing the UCDs in the first epoch, and seven targets still have potential to harbour faint companions but did not get observed in the second epoch, there are $n = 66-2-7 =57$ independent trials. We have
\begin{equation}
    (1 - p_{\mathrm{UCD}})^{57} \le \alpha = 1 - 90\%,
\end{equation}
yielding a true frequency with an upper limit of $p_{\mathrm{UCD}}\le 4.0\%$. 
This frequency limit only applies to the complete sample with spectral types earlier than esdT0.

%
%
\section{Discussion}
\label{discussion}

\begin{table*}
\centering
\caption{Summary of companion frequencies from this work and the literature.}
\label{tab:freq_summary}
\resizebox{\linewidth}{!}{%
\renewcommand{\arraystretch}{1.4}
\begin{tabular}{p{2.4cm} p{1.6cm} p{1.6cm} p{2.85cm} p{1.5cm} p{6.2cm}}
\hline\hline
Primary & Secondary & [Fe/H] & Projected separation & Frequency & Reference(s) \\
 (Sp. T.) &  & (dex) & (au) & (\%) &  \\
\hline
FGK$^{(\mathrm{a})}$ & UCD & $<-1.5$ & $\sim10^2$--$\sim10^4$ & $<4.0$ & This work \\
\hline
Young FGK & BD & $\sim0$ & 28--1590  & $3.2^{+3.1}_{-2.7}$ & \citet{metchev2009companion_mass_func}\\
USco young star & UCD cand. & $\sim0$ & 400--9000 & $\sim3$ & \citet{chinchilla2019companion_young}\\
Mixed star & UCD cand. & ... & 1000--24,000  & 2--4 & \citet{dalponte2020ultracool_binary} \\
\hline
M & UCD & $\sim-1.0$ & 1000--140,000  & $1.9^{+3.7}_{-1.9}$ & \citet{gonzalez-payo2021_wide_companion_sdML} \\
M & UCD & $\sim-2.0$ & 1000--140,000 & $<5.3$ & \citet{gonzalez-payo2021_wide_companion_sdML}\\
M & BD & $\sim0$ & Few--7500 & $1.3\pm0.3$ & \citet{winters2019multiplicity_dM_25pc} \\
\hline
FGK& UCD & $<-1.5$ & $<10$ & $<$ Few & This work, inferred from other studies\\
\hline
FGK & Star & $<-1.5$ & $\sim10^2$--$\sim10^4$ & $6.1^{+7.2}_{-4.0}$ & This work \\
FGK & Star & $<-1.6$ & $>10$ ($>300$) & 20 ($\sim9$) & \citet{zinnecker2004halostar_binary}, $K$ flux ratio $>0.01$ \\
FGK & Star & $<-1.5$ & 8--10,000 & $<3$ & \citet{lodieu2025widebinary} \\
FG & Star & $\sim-1.5$ & 1000--10,000 & $\sim1$ & \citet{hwang2021binary_metallicity} \\
GKM & Star & $<-1.5$ & 32--57,000 & 4.5 & \citet{lodieu2025widebinary,zapatero2004metalpoor_wide_binary} \\
KM & Star & $<-0.5$ & $>100$ & $>2.41$ & \citet{zhang2013SDSS_sdM_binary} \\
KM & Star &  $<-0.5$ & $>100$ & 14 & \citet{jao2009cool_subdwarf_companion_speckle} \\
FGKM & Star & metal-poor & 100--100,000 & $12.5\pm1.9$ & \citet{ziegler2015cool_subdwarf_companion_highres} \\
\hline
M & UCD & metal-poor &  Few--tens  & $\sim3$ & \citet{riaz2008sdM_binary,lodieu2009sdM_binary}\\
M & BD & ... & $<2$  & 0.3  & \citet{gaia2023dr3_multiplicity} \\
M & BD & ... & Few--tens & 2.3--2.8  & \citet{dietrich2012multiplicity_5_70au,bowler2015dM_BD,susemiehl2022dM_binary} \\
\hline
\end{tabular}
}
\begin{tablenotes}[flushleft]
\item \textbf{Note:} Groups with different physical properties are separated by horizontal lines. Values in parentheses are derived from the literature data.
\item $^{(\mathrm{a})}$: Some of the targets have ambiguous spectral type classification or earlier classification from the literature, such as HD\,340279, which was classified as early as A5 by \citet{bidelman1985kuiper_pm} and then as F8 by \citet{nesterov1995hd}. According to our targets' effective temperatures from 4600 to 6300\,K, we generalized their types as FGK.
\end{tablenotes}
\end{table*}

We compared our results with literature samples spanning a range of physical properties, in order to explore potential dependencies of the wide ultracool companion frequency. A summary of this discussion is provided in Table~\ref{tab:freq_summary}.

\subsection{Metallicity and age dependence}
Our $p_{\mathrm{UCD}}$ is in line with the 2$\sigma$-limit frequency of wide (28--1590\,au) brown dwarf (BD) companions to young solar-type stars of $3.2^{+3.1}_{-2.7}\%$ \citep{metchev2009companion_mass_func}. Similarly, in Upper Scorpius, \citet{chinchilla2019companion_young} found a comparable $\sim$3\% frequency of wide (400--9000\,au) companions with candidate spectral types M to L around young stars. Our result is also consistent with the wide (1000--24,000\,au) star-UCD candidate binary fraction of 2--4\% \citep{dalponte2020ultracool_binary}. 

These consistencies do not reveal the metallicity or age influence on the formation of the binaries consisting of a solar-type star and an ultracool companion. Their pronounced paucities stay universal across different metallicities and ages. 

\subsection{Primary mass dependence}
Our $p_{\mathrm{UCD}}$ aligns with the wide-companion frequency to metal-poor late-M and L subdwarfs reported by \citet{gonzalez-payo2021_wide_companion_sdML}, who measured a multiplicity rate of $1.0^{+2.0}_{-1.0}\%$ for M subdwarfs, $1.9^{+3.7}_{-1.9}\%$ for extreme M subdwarfs, and set an upper limit of $5.3\%$ for ultra M subdwarfs, although over projected separations of $\sim$ 1000--140,000\,au. Since their sample included only one companion to an esdM --- which is an esdL --- and no companion to usdMs, the multiplicity rates for esdMs and usdMs effectively represent their ultracool companion frequencies or limits. 
As a result, there is no evidence showing that the low ultracool companion frequency in the metal-poor regime is dependent on the primary mass. 

Regardless of the metallicity, \citet{winters2019multiplicity_dM_25pc} revealed that for nearby M dwarfs within 25\,pc, the wide BD companion frequency is $1.3\pm0.3\%$ with separation between 2 and 300\,arcsec, which is equivalent to from a few au to a maximum 7500\,au.  
The consistency between this value and our result supports the conclusion that, at present, there is no clear evidence that low metallicity influences the wide ultracool companion frequency, even across different primary stellar types --- from FGK stars to M dwarfs.

\subsection{Primary-secondary separation dependence}

Although there has been no frequency of the close ultracool companion to metal-poor stars reported so far, we can infer it from other studies. \citet{marcy2000BD_solartype_star} and \citet{grether2006BD_desert} found < 1\% solar-type stars harbour BDs in close orbits, so-called the BD desert. \citet{sahlmann2011BD_solar_companion} set the upper limit of the frequency of close ($<$\,10\,au) BD companions around solar-metallicity solar-type stars to 0.6\%. 

According to \citet{moe2019closebinary}, the close binary rate of solar-type stars has strong anti-correlation with the metallicity of the primary: it increases from $24\pm6\%$ at a field metallicity to $53\pm12\%$ at a metallicity of $-3$\,dex by a factor of two. Close companions form mostly via disk fragmentation, and this instability is more likely to occur under lower metallicity conditions due to higher infall rates from hotter cores and lower temperatures of the optically thick disks. Even if we take a bold assumption that close BD companion frequency follows this trend with metallicity, the frequency of close BD companions to metal-poor stars will not be superior than a few percent, which is still in line with our result of wide ultracool frequency.

Giant planets serve as analogues to UCDs, although they may form in a different manner than more massive UCDs near a metal-poor star. \citet{mortier2012metalpoor_star_giant_planets} claimed that giant planets are rare ($\leq2.36\%$) around metal-poor stars with metallicity [Fe/H] < $-0.7$\,dex, which is comparable to our inferred low frequency for close BD companions to metal-poor stars. However, if the fact that close BD companions appears rarer than giant planets \citep{grether2006BD_desert} still holds for metal-poor stars, we would expect a even lower frequency for the close ultracool companion frequency. Overall, these comparisons suggest that separation may not strongly influence the ultracool companion frequency around metal-poor stars.

\subsection{Stellar companion frequency}

Although previous studies report varying results on stellar companion frequencies for metal-poor stars, our $p_{\mathrm{star}}$ is consistent with them within the uncertainties and considering incompleteness effects. 
Using infrared speckle interferometry on the same sample of \citet{carney1994haloFGK}, together with that of \citet{norris1986halo}, \citet{zinnecker2004halostar_binary} derived a binary frequency of 6--20\% for halo stars with metallicities [Fe/H] $<-1.6$\,dex and separations larger than 10\,au. The 6\% estimate corresponds to a $K$-band flux-ratio detection threshold of 0.1, while the 20\% estimate assumes a threshold of 0.01. To enable a direct comparison with our result, we excluded companions within 300\,au from their background-corrected sample at 0.01 $K$-band flux ratio threshold, yielding a frequency of $\sim9\%$ for binaries wider than 300\,au. It is an approximation since there are triple systems included. Our result is consistent with this value.
\citet{lodieu2025widebinary} reported an upper limit of 3\% for stellar companions to mid-F- to early-K primaries with metallicity [Fe/H] $<-1.5$,dex and projected separations between 8 and 10,000\,au. Although this value is relatively low and may be strongly affected by incompleteness, the lower bound of our result is consistent with this limit.
\citet{hwang2021binary_metallicity} reported a relatively low wide (1000 to 10,000\,au) binary fraction for metal-poor FG stars with metallicity $\sim-1.5$\,dex of $\sim1\%$ using LAMOST data, which is likewise consistent with our findings considering the uncovered gap between 100 to 1000\,au.
Our result is further in line with the 4.5\% frequency reported by \citet{lodieu2025widebinary} after correcting for the primaries’ metallicities, based on the overall wide-binary fraction of 13--15\% for stars with metallicities between $-3.5$ and 0.0\,dex and projected separations between 32 and 57,000\,au from \citet{zapatero2004metalpoor_wide_binary}. Our result also agrees with the lower bound of the wide ($>100$\,au) binary frequency of 2.41\% for KM red subdwarfs across different metallicity subclasses reported by \citet{zhang2013SDSS_sdM_binary}. \citet{jao2009cool_subdwarf_companion_speckle} provided a high multiplicity rate of metal-poor KM-type cool subdwarfs of $26\pm6$\% using speckle interferometry. Within the 26\%, 12\% is for binaries with separation less than 100\,au the rest 14\% for wide binaries farther than 100\,au. 
\citet{ziegler2015cool_subdwarf_companion_highres} probed wide companions to metal-poor FGKM-type subdwarfs using high-resolution adaptive optics imaging and included previously recorded wide companions. They found a multiplicity rate of $12.5\pm1.9\%$, and those systems have projected separations ranging from $105\pm12$ to $79,156 \pm 9046$\,au.  At its upper bound, our wide companion frequency is consistent with both studies. Two factors may account for this agreement. First, similar to \citet{zinnecker2004halostar_binary}, these high-resolution studies effectively covered the separation range from 100 to a few hundred au that were not probed by our seeing-limited observations. Second, they targeted lower-mass primaries that were not studied by this work.

Following the review of wide stellar companion frequencies reported in the literature, we adopted here only our own measurement $p_{\mathrm{star}}$ for metal-poor halo stars. Based on this reference, the wide ultracool companion frequency $p_{\mathrm{UCD}}$ is marginally lower than that of wide stellar companions $p_{\mathrm{star}}$.

\subsection{Primary mass and separation}

As discussed above that primary-secondary separation may have little effect on the ultracool companion frequency for metal-poor FGK stars.  In the metal-poor M subdwarf regime, the wide companion frequency of $1.0^{+2.0}_{-1.0}\%$ \citep{gonzalez-payo2021_wide_companion_sdML} is comparable with the frequency of intermediate-separated (a few to tens of au) companions to M subdwarfs of $\sim3\%$ measured by \citet{riaz2008sdM_binary} and \citet{lodieu2009sdM_binary} using the \textit{Hubble} Space Telescope and lucky imaging, respectively. 

In addition, although lacking metallicity constraints but with tighter limits on secondary masses, \citet{gaia2023dr3_multiplicity} proposed $\sim0.3\%$ for the frequency of close BDs to M dwarfs with periods less than about 1000 days, i.e., with semi-major axes less than 1 to 2\,au. Several studies also agree to give low frequencies of BD companions to M dwarfs of 2.3\% to 2.8\% at intermediate separations of a few to tens of au \citep{dietrich2012multiplicity_5_70au,bowler2015dM_BD,susemiehl2022dM_binary}. Taken together, these results suggest that separation may have little effect on the ultracool companion frequency around metal-poor stars across all spectral types, and around M dwarfs across a range of metallicities.

\subsection{Rarity of wide ultracool companions}

We found that the wide ultracool companions are rare around stars. This scarcity appears consistent across primary spectral types, metallicities, and companion separations. 
The formation of wide ultracool companions may be naturally suppressed in core fragmentation, as the efficiency of fragmentation decreases toward the low-mass end \citep{chabrier2003IMF,bate2012formation_simulation}, and extreme mass-ratio ($q<0.1$) systems represent only a small fraction of the outcomes of the formation process \citep{bate2012formation_simulation}.

In particular, even if metal-poor environment favours the formation of wide ultracool companions, they have gone through a long cooling process of $\sim10$\,Gyr. The degeneracy of BDs and very-low-mass stars in the spectral type range late-M to L naturally breaks: the majority of substellar objects have been cooled down to very late spectral types \citep{zhangzenghua2019metal-poor_sdT}, which are beyond the detection limit of this research, leaving a spectral type gap from early-L to early-T barely filled by transitional BDs. These transitional BDs occupy a very narrow mass range and thus account for a small portion of the population \citep{zhang2017usdL, zhangzenghua2018transitionalBD,zhangzenghua2019metal-poor_sdT}.

In addition, halo wide binaries may not be stable enough to survive the interstellar interactions through their lifetime. For binaries with total masses of 1 M$_\odot$ and semi-major axes of 1000 and 10,000 au, the simulation yields $\sim90\%$ and $\sim20\%$ probabilities of surviving for 10\,Gyr, respectively \citep{weinberg1987binary_fate}. For our star-UCD systems, they could be more easily disrupted at lower total binding energies due to high-mass-ratio at a certain total mass.

\section{Conclusion}
\label{conclusion}

We did not identify any bona fide ultracool comoving companion to all 57 halo stars. We concluded that the wide ultracool companion frequency $p_{\mathrm{UCD}}$ with companion spectral types earlier than esdT0 around halo metal-poor stars, within the range of separation of typically a few hundred au up to a few thousand au, is less than 4.0\% at a 90\% confidence level.

This frequency is marginally lower than the wide stellar companion frequency around halo metal-poor stars, for which we found using the total 66 samples $p_{\mathrm{star}}=6.1^{+7.2}_{-4.0}\%$. Given the current uncertainties, we found no statistically significant evidence for any dependence of the wide ultracool companion frequency on metallicity, separation, and primary mass. The ultracool companion appears rare across all explored physical parameter space.

We speculated that, for the halo population, most brown dwarfs could have cooled to very late spectral types and thus fall below our detection limits of $\sim$\,esdT0. The resulting low upper limit of $p_{\mathrm{UCD}}$ may therefore reflect only the small fraction of transitional brown dwarfs, which occupy a narrow mass range. More generally, ultracool companions may be intrinsically disfavoured in formation, and wide binary systems in the halo may have a reduced survival probability over their long lifetimes.

Although seven of our targets with potential faint companions currently lack a second epoch, the very low upper limit on the frequency of wide ultracool companions to halo stars implies that a survey-driven strategy is more effective than continued deep NIR imaging of individual targets, even with a 10-m class telescope. In this context, combining ongoing and forthcoming deep NIR surveys such as \textit{Euclid} \citep{zhangjerry2024euclid_reconnaissance,zerjal2025ucd,mohandasan2025ucd} and \textit{Roman} \citep{holwerda2023roman_ucd} offers a substantially more efficient approach.

\begin{acknowledgements}
We thank our referee, Prof. ZengHua Zhang for providing insightful comments and suggestions to this work. 
Funding for this research was provided by the Agencia Estatal de Investigación del Ministerio de Ciencia e Innovación (AEI-MCINN) under grants PID2019-109522GB-C53 and PID2022-137241NB-C41 as well as the European Union (ERC, SUBSTELLAR, project number 101054354). 
JYZ also thanks the support from the Western Postdoctoral Fellowship provided by Western University.
Based on observations made with the Gran Telescopio Canarias (GTC), in the Spanish Observatorio del Roque de los Muchachos of the Instituto de Astrofísica de Canarias, on the island of La Palma, under programmes GTC53-20B, GTC65-21A, and GTC45-24A (PI Lodieu).
EMIR has been funded by GRANTECAN S.L.\ via a procurement contract; by the Spanish funding agency grants AYA2001-1656, AYA2002-10256-E, FIT-020100-2003-587, AYA2003-01186, AYA2006-15698-C02-01, AYA2009-06972, AYA2012-33211, AYA2015-63650-P and AYA2015-70498-C2-1-R; and by the Canarian funding agency grant ACIISI-PI 2008/226.
This research has made use of the Spanish Virtual Observatory (https://svo.cab.inta-csic.es) project funded by MCIN/AEI/10.13039/501100011033/ through grant PID2020-112949GB-I00. 
This research has made use of the Simbad and Vizier databases, operated at the centre de Donn\'ees Astronomiques de Strasbourg (CDS), and of NASA's Astrophysics Data System Bibliographic Services (ADS).
This work made use of Astropy: a community-developed core Python package and an ecosystem of tools and resources for astronomy \citep[\url{http://www.astropy.org}; ][]{astropy2013, astropy2018, astropy2022}.
This research has made use of the Washington Double Star Catalog maintained at the U.S. Naval Observatory.
\end{acknowledgements}

%
%

%
\bibliographystyle{aa} 
\bibliography{bibliography.bib} 

\onecolumn

\begingroup
\setlength{\tabcolsep}{3pt} 

\fontsize{9}{10}\selectfont

\begin{appendix}
\begin{onecolumn}
\section{Target information}

\begin{longtable}{lccccccccccc}
\caption{Observed halo metal-poor stars.}\\  
\hline\hline
Name & $\alpha$ & $\delta$ & $\mu_{\alpha}\cos{\delta}$ & $\mu_{\delta}$ & $d$ & $T_{\mathrm{eff}}$ & $J$ & [Fe/H] & $t_1$ & $t_2$ & $\Delta t$ \\
& (\si{deg}) & (\si{deg}) & (\si{mas.yr^{-1}}) & (\si{mas.yr^{-1}}) & (\si{pc}) & (K) & (\si{mag}) & (dex) & (\si{MJD}) & (\si{MJD}) & (\si{yr}) \\
\hline
\endfirsthead
\caption{Stellar Proper Motion Data (continued)}\\
\hline\hline
Name & $\alpha$ & $\delta$ & $\mu_{\alpha^*}$ & $\mu_{\delta}$ & Dist. & $T_{\mathrm{eff}}$ & $J$ & [Fe/H] & $t_1$ & $t_2$ & $\Delta t$ \\
& (\si{deg}) & (\si{deg}) & (\si{mas.yr^{-1}}) & (\si{mas.yr^{-1}}) & (\si{pc}) & (K) & (\si{mag}) & (dex) & (\si{MJD}) & (\si{MJD}) & (\si{yr}) \\
\hline
\endhead
\hline
\multicolumn{8}{r}{{Continued on next page}} \\
\endfoot
\hline\hline
\endlastfoot

Ross  541 & 2.0819179 & $-5.2485839$ & 352.9 & $-131.3$ & 193.9 & 5341 & 10.9 & $-1.63$ & 59788.21 & ... & ... \\
Ross  312 & 10.0954937 & 7.3750678 & 264.4 & $-122.1$ & 182.3 & 4858 & 11.6 & $-1.58$ & 59788.19 & ... & ... \\
BD+71    31 & 10.9347892 & 72.1786447 & 324.0 & 92.2 & 180.4 & 6082 & 9.2 & $-2.27$ & 59148.96 & 60513.18 & 3.74 \\
BD+15   150 & 15.2770967 & 16.3726544 & 340.6 & $-149.3$ & 160.1 & 5631 & 9.4 & $-1.71$ & 59148.92 & ... & ... \\
BD+47   435 & 22.8162862 & 48.0048403 & 311.2 & $-36.6$ & 162.2 & 6061 & 9.2 & $-1.94$ & 59148.94 & 60510.16 & 3.73 \\
G  34-45 & 25.7032042 & 22.6171922 & 74.1 & $-301.6$ & 141.4 & 4697 & 11.3 & $-1.54$ & 59243.89 & ... & ... \\
G  72-34 & 26.5153625 & 35.9134931 & 91.5 & $-381.5$ & 143.9 & 4689 & 11.3 & $-2.21$ & 59243.87 & ... & ... \\
Ross   18 & 33.7895458 & 32.3949656 & 455.2 & $-152.1$ & 129.4 & 4858 & 10.9 & $-1.78$ & 59239.86 & ... & ... \\
HD  16031 & 38.5460354 & $-12.3842900$ & 60.1 & $-184.3$ & 121.5 & 6031 & 8.8 & $-1.82$ & 59123.08 & ... & ... \\
BD+09   352 & 40.3068317 & 9.7700278 & 305.1 & $-15.7$ & 158.9 & 5969 & 9.0 & $-2.28$ & 59121.20 & ... & ... \\
G   4-36 & 40.8418679 & 13.4324947 & 339.3 & $-147.2$ & 190.0 & 5917 & 10.4 & $-2.12$ & 59243.91 & ... & ... \\
G   5-19 & 47.8605858 & 12.6193250 & $-24.7$ & $-467.6$ & 160.0 & 5583 & 9.8 & $-1.69$ & 59178.06 & $^{(\mathrm{c})}$ & ... \\
G  79-56$^{(\mathrm{a})}$ & 55.4267775 & 9.3925450 & 262.0 & $-251.2$ & 128.7 & 5184 & 10.3 & $-1.5$ & 59239.89 & ... & ... \\
G 246-49 & 57.6032667 & 64.7844639 & 311.3 & $-397.4$ & 167.3 & 4984 & 11.5 & $-1.58$ & 59123.10 & ... & ... \\
HD  25329 & 60.8124912 & 35.2732792 & 1731.6 & $-1364.4$ & 18.5 & 4762 & 6.8 & $-1.73$ & 59121.22 & ... & ... \\
BD+03   740 & 75.3192608 & 4.1102836 & 155.3 & $-146.3$ & 169.7 & 6145 & 8.8 & $-2.78$ & 59121.25 & $^{(\mathrm{c})}$ & ... \\
BD+58   876 & 89.3690979 & 58.6801867 & 192.2 & $-108.4$ & 130.2 & 5835 & 9.4 & $-1.94$ & 59123.24 & ... & ... \\
G  99-48 & 89.7748950 & 4.1773844 & 264.9 & $-231.4$ & 146.9 & 5076 & 10.4 & $-2.14$ & 59187.08 & $^{(\mathrm{c})}$ & ... \\
G 103-28 & 94.4049912 & 28.6471153 & 161.3 & $-249.6$ & 133.6 & 4676 & 11.3 & $-1.6$ & 59178.08 & $^{(\mathrm{c})}$ & ... \\
G 103-50 & 100.0394179 & 28.4502489 & 256.0 & $-259.1$ & 139.6 & 4642 & 10.3 & $-2.0$ & 59239.93$^{(\mathrm{b})}$ & ... & ... \\
BD+00 2058A$^{(\mathrm{a})}$ & 115.9331946 & $-0.0669369$ & $-171.9$ & $-305.5$ & 178.7 & 5961 & 9.2 & $-1.27$ & 59188.17 & ... & ... \\
HD  64090 & 118.3880033 & 30.6050711 & 706.7 & $-1835.4$ & 27.3 & 5381 & 7.0 & $-1.75$ & 59213.04 & ... & ... \\
G 234-24 & 122.5701337 & 69.7812186 & 279.1 & $-77.2$ & 146.6 & 5900 & 9.9 & $-1.6$ & 59239.94 & ... & ... \\
HD 233511 & 124.8440475 & 54.0860064 & $-34.0$ & $-628.3$ & 98.3 & 5968 & 8.6 & $-1.64$ & 59187.11 & ... & ... \\
HD  74000 & 130.2116817 & $-16.3451433$ & 351.0 & $-484.0$ & 110.1 & 6141 & 8.7 & $-2.02$ & 59267.89 & ... & ... \\
G 115-49 & 136.3195133 & 38.7984983 & 6.5 & $-468.5$ & 165.6 & 5694 & 10.5 & $-2.19$ & 59239.97 & ... & ... \\
Ross  889 & 145.1800221 & 1.0081975 & 148.0 & $-505.5$ & 180.0 & 6295 & 9.6 & $-2.66$ & 59268.96 & $^{(\mathrm{c})}$ & ... \\
Ross  890 & 147.4649767 & 6.6098656 & 74.3 & $-326.6$ & 168.9 & 5269 & 11.1 & $-2.28$ & 59269.00 & ... & ... \\
Wolf  334 & 149.4264854 & 32.6131328 & $-373.5$ & $-271.9$ & 178.1 & 5084 & 11.3 & $-1.82$ & 59239.99 & ... & ... \\
G 236-30 & 157.2015667 & 62.9957200 & 183.4 & $-243.0$ & 235.0 & 5311 & 11.4 & $-2.22$ & 59268.98 & ... & ... \\
G  10-54 & 177.4507842 & 6.1478558 & 32.8 & $-315.6$ & 188.9 & 5315 & 11.2 & $-2.05$ & 59240.14 & ... & ... \\
BD+26  2606 & 222.2598304 & 25.7025533 & $-9.7$ & $-351.2$ & 147.6 & 5973 & 8.7 & $-2.58$ & 59448.87 & ... & ... \\
BD+42  2667 & 240.8054117 & 42.2462919 & $-195.6$ & $-365.2$ & 117.8 & 5973 & 8.8 & $-1.53$ & 59276.27 & 60186.92 & 2.49 \\
G 168-26 & 240.8326625 & 21.9698989 & $-299.8$ & $-93.7$ & 135.5 & 5582 & 9.9 & $-1.8$ & 59360.08 & ... & ... \\
G 153-64 & 248.1251300 & $-8.5606200$ & $-147.7$ & $-214.3$ & 158.6 & 5221 & 9.9 & $-1.54$ & 59414.98 & $^{(\mathrm{c})}$ & ... \\
G 169-28 & 252.5477854 & 22.3138950 & $-149.4$ & $-366.6$ & 153.7 & 5597 & 10.0 & $-1.6$ & 59360.10 & ... & ... \\
G 138-65 & 252.7729775 & 15.8635486 & $-269.8$ & $-332.3$ & 183.2 & 4756 & 12.0 & $-1.88$ & 59364.14 & 60449.99 & 2.97 \\
G 181-19 & 254.7894767 & 34.8657178 & $-212.3$ & $-163.9$ & 138.1 & 4872 & 11.0 & $-1.56$ & 59360.18 & ... & ... \\
G 139-8 & 255.4332558 & 16.1509283 & $-287.5$ & $-246.4$ & 190.5 & 5909 & 10.3 & $-2.63$ & 59360.12 & ... & ... \\
G 227-10 & 264.0249033 & 63.5672847 & $-246.1$ & 100.5 & 156.8 & 5175 & 11.0 & $-1.59$ & 59394.09 & ... & ... \\
BD+02  3375 & 264.9399783 & 2.4165572 & $-365.9$ & 75.2 & 101.6 & 5852 & 8.8 & $-2.55$ & 59393.09 & 60450.09 & 2.89 \\
BD-08 4501 & 266.8665467 & $-8.7799258$ & 246.4 & $-364.3$ & 114.7 & 5657 & 9.2 & $-1.99$ & 59415.03 & $^{(\mathrm{c})}$ & ... \\
  &  &  &   &   &  &   &   &   &  59419.05 &  &  \\
BD+36  2964 & 268.0752187 & 36.4018386 & $-155.8$ & $-244.5$ & 166.7 & 5952 & 9.4 & $-2.53$ & 59364.16 & 60514.10 & 3.15 \\
BD+20  3603 & 268.6801275 & 20.2712344 & $-223.8$ & $-349.8$ & 123.9 & 6200 & 8.9 & $-2.26$ & 59360.21 & 60450.03 & 2.98 \\
G 206-1 & 269.9268217 & 36.3929569 & $-200.0$ & $-59.7$ & 136.7 & 4987 & 10.9 & $-1.68$ & 59365.16 & ... & ... \\
HD 188510 & 298.7903258 & 10.7409439 & $-38.4$ & 290.6 & 38.1 & 5450 & 7.6 & $-1.75$ & 59360.14 & 60479.05 & 3.06 \\
G  24-3 & 301.4346542 & 4.0480033 & $-133.7$ & $-167.8$ & 130.7 & 5865 & 9.4 & $-1.81$ & 59122.90 & ... & ... \\
HD 340279 & 306.1892187 & 25.0519747 & 89.3 & $-247.8$ & 197.9 & 6231 & 9.9 & $-2.85$ & 59096.02 &  &  \\
  &   &   &  &   &   &   &   &  & 59393.21 & 60479.00 & 2.97 \\
BD+41  3931 & 313.8198129 & 42.3001889 & 55.4 & $-390.0$ & 71.6 & 5445 & 9.0 & $-1.78$ & 59122.87 & 60479.02 & 3.71 \\
G  25-24 & 319.1734712 & $-1.3024353$ & 266.3 & $-249.0$ & 180.5 & 5774 & 10.5 & $-1.94$ & 59363.20 & 60513.10 & 3.15 \\
BD+26  4136 & 320.4906083 & 27.4528847 & 205.0 & 153.3 & 106.2 & 5734 & 9.4 & $-1.55$ & 59362.17 & 60450.21 & 2.98 \\
G 214-1$^{(\mathrm{a})}$ & 326.9817604 & 33.1075447 & 195.7 & $-14.4$ & 179.0 & 5568 & 10.8 & $-2.03$ & 59177.86 & ... & ... \\
G 214-5 & 329.7932142 & 41.0414150 & $-292.2$ & $-192.3$ & 162.7 & 5692 & 10.4 & $-2.12$ & 59362.19 & 60450.18 & 2.98 \\
Wolf 1337 & 329.9785792 & $-0.6630075$ & $-175.8$ & $-326.6$ & 190.7 & 4976 & 11.6 & $-1.67$ & 59419.07 & ... & ... \\
G  27-8$^{(\mathrm{a})}$ & 330.8072077 & $-1.2208872$ & 198.0 & $-127.6$ & 172.6 & 5844 & 10.2 & $-1.53$ & 59419.09$^{(\mathrm{b})}$ & ... & ... \\
G 241-4 & 335.3389850 & 68.4637908 & $-261.0$ & 143.7 & 158.5 & 5132 & 11.4 & $-1.59$ & 59394.12 & 60186.95 & 
\\
&  &  & &  &  &  &  &  &  & 60479.13 & 2.97 \\
L 1077-30 & 337.9009096 & 2.1621800 & 50.0 & $-325.4$ & 190.2 & 5861 & 9.6 & $-1.75$ & 59395.18 & ... & ... \\
G 189-50 & 344.1144946 & 33.8843306 & $-98.8$ & $-371.4$ & 192.5 & 5259 & 11.3 & $-1.59$ & 59362.21 & 60514.12 & 3.15 \\
Ross  783 & 346.9989929 & 41.8556644 & 345.8 & $-93.1$ & 133.9 & 5330 & 9.9 & $-1.92$ & 59095.07 & 60513.12 & 3.88 \\
G  28-42 & 347.3414187 & 7.0205764 & 377.1 & $-19.4$ & 172.1 & 5363 & 10.9 & $-1.71$ & 59448.98 & 60514.15 & 2.92 \\
BD-00  4470 & 347.3872679 & 0.7110414 & $-221.9$ & $-1295.2$ & 73.0 & 5072 & 8.5 & $-1.8$ & 59782.21 & 60479.15 & 1.91 \\
BD+38  4955 & 348.4117521 & 39.4173858 & 175.7 & $-315.2$ & 84.1 & 5127 & 9.5 & $-2.68$ & 59096.05 &  &  \\
  &   &   &   &   &   &   &   &   & 59393.11 & 60479.18 & 2.97 \\
Ross  304 & 355.3520592 & 59.4096989 & 50.3 & $-222.2$ & 147.5 & 5046 & 11.1 & $-2.17$ & 59452.14 & 60513.16 & 2.90 \\
G 130-7 & 356.2502137 & 30.3362225 & 212.3 & $-179.2$ & 130.7 & 5306 & 10.3 & $-1.62$ & 59395.20 & ... & ... \\
G 171-15 & 356.2612783 & 44.6676194 & 48.0 & $-235.4$ & 115.7 & 5309 & 10.2 & $-2.12$ & 59096.07 & 60513.14 & 3.88 \\
Wolf 1043 & 357.5056904 & 8.7231586 & 369.3 & $-55.2$ & 127.3 & 5643 & 10.1 & $-2.42$ & 59095.09 & ... & ... 
\label{targets}
\end{longtable}
\textbf{Notes:} A total of 66 targets were observed during the first epoch ($t_1$). Among these, 28 were selected for second-epoch ($t_2$) follow-up observations, with 21 successfully observed at the end.

$^{(\mathrm{a})}$: The four targets that have wide stellar companions confirmed by \textit{Gaia}.

$^{(\mathrm{b})}$: The two targets that had depth issues in the first epoch.

$^{(\mathrm{c})}$: The seven targets that remained unobserved in the second epoch.

\vfill

\section{Dual epoch images of the 21 targets}
\begin{figure}[htbp] 
    \centering
    \includegraphics[width=0.8\textwidth]{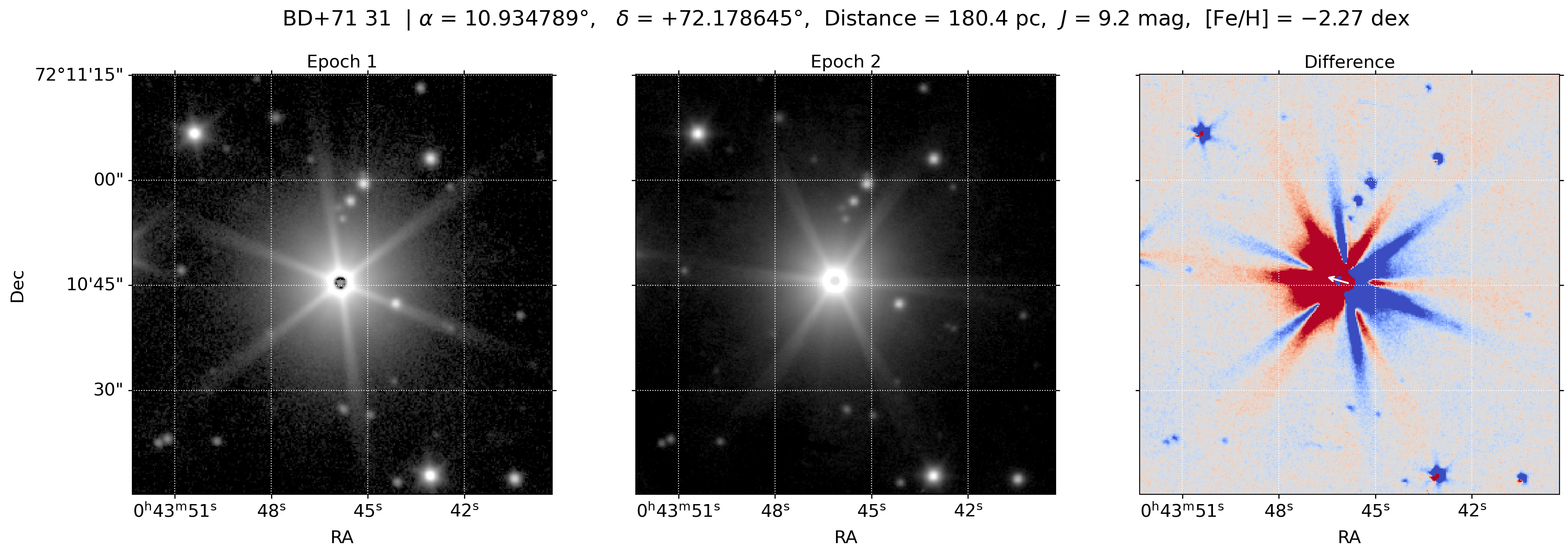}
    \includegraphics[width=0.8\textwidth]{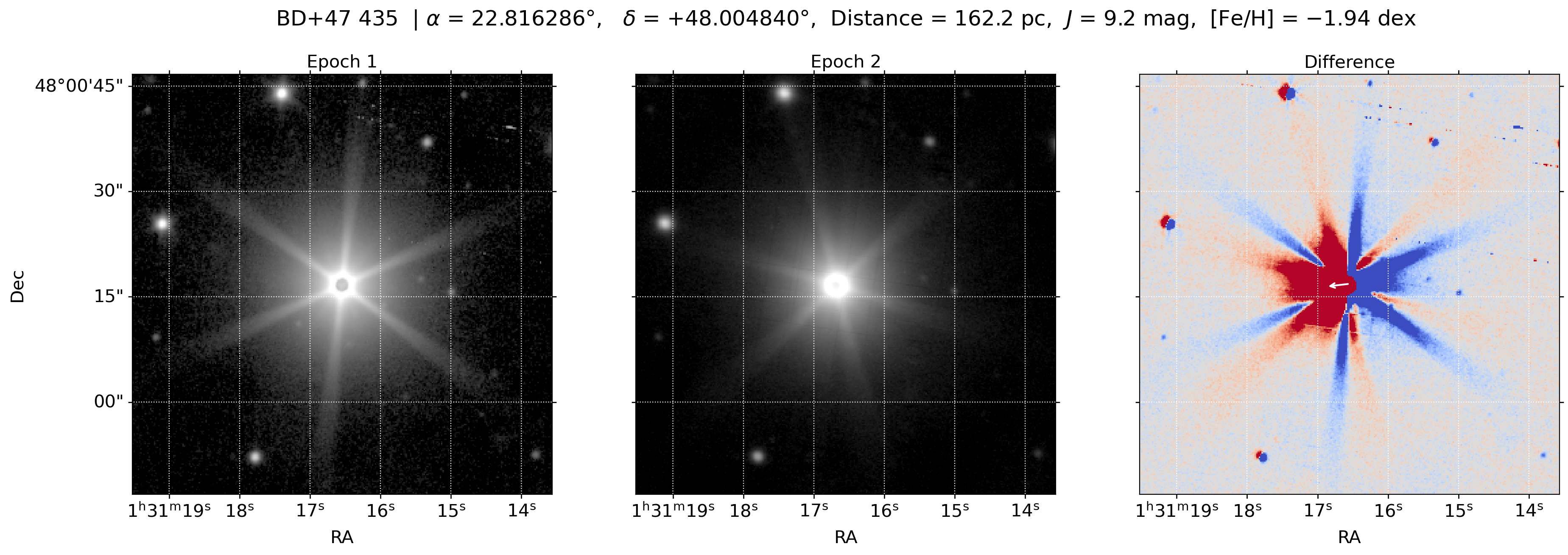}
    \caption{GTC/EMIR $J$-band images of the 21 targets in two epochs and the difference (red positive, blue negative). The two-epoch images are normalised and then stretched on a logarithmic scale to better visualise the faint sources around the central star. White arrow in the difference image indicates the proper motion direction of the star. For clarity, the arrow length is three times the motion during the baseline between the two epochs. All images are centred at the star position at the first epoch with an FoV of 1\arcmin $\times$ 1\arcmin.}
    \label{dual}
\end{figure}

\newpage
Figure~\ref{dual} continues:

\begin{figure}[htbp]
\centering
    \includegraphics[width=0.8\textwidth]{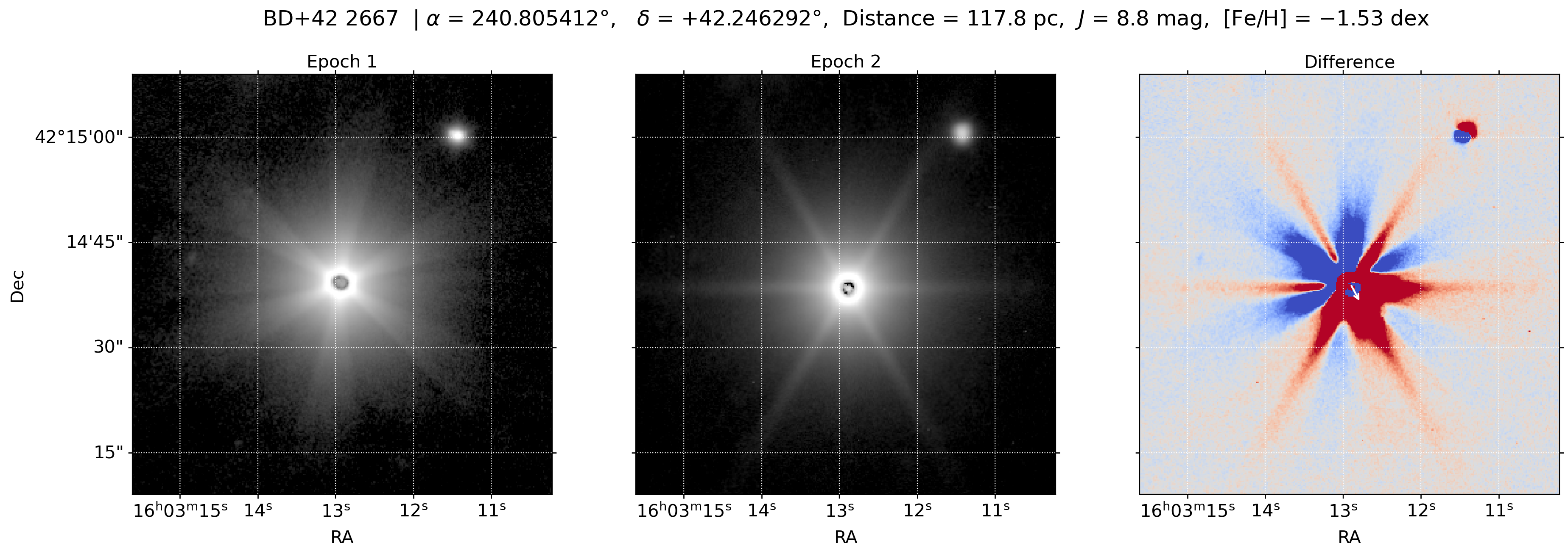}
    \includegraphics[width=0.8\textwidth]{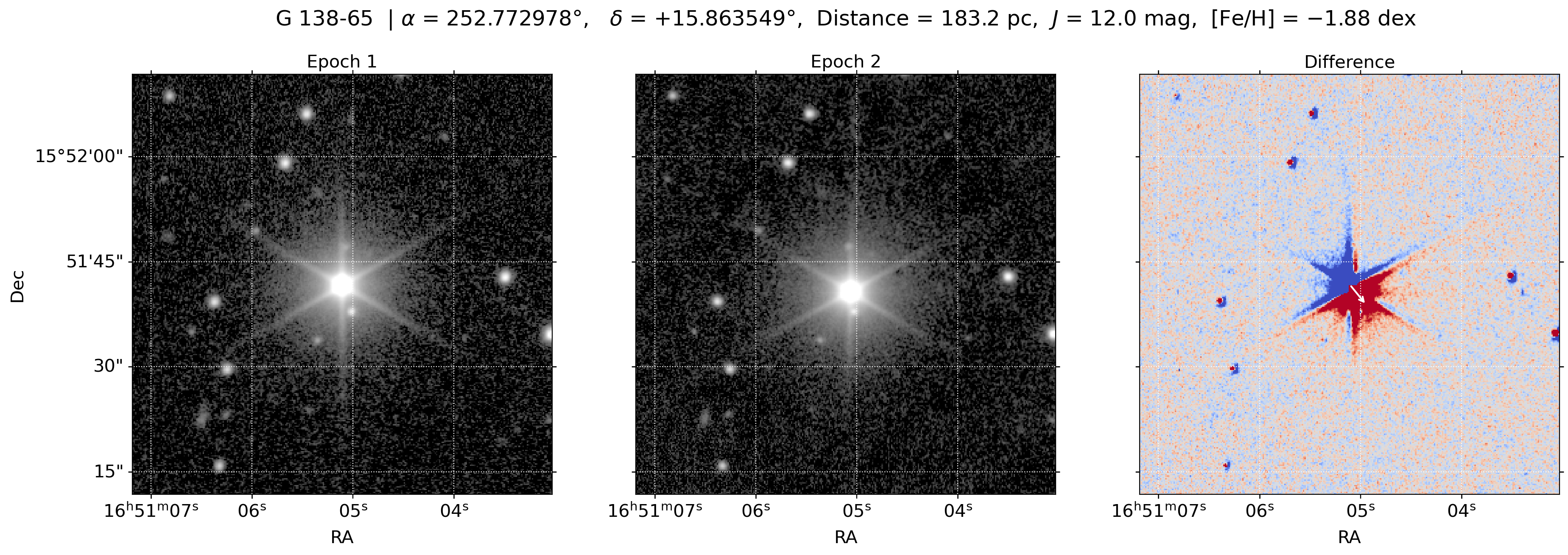}
    \includegraphics[width=0.8\textwidth]{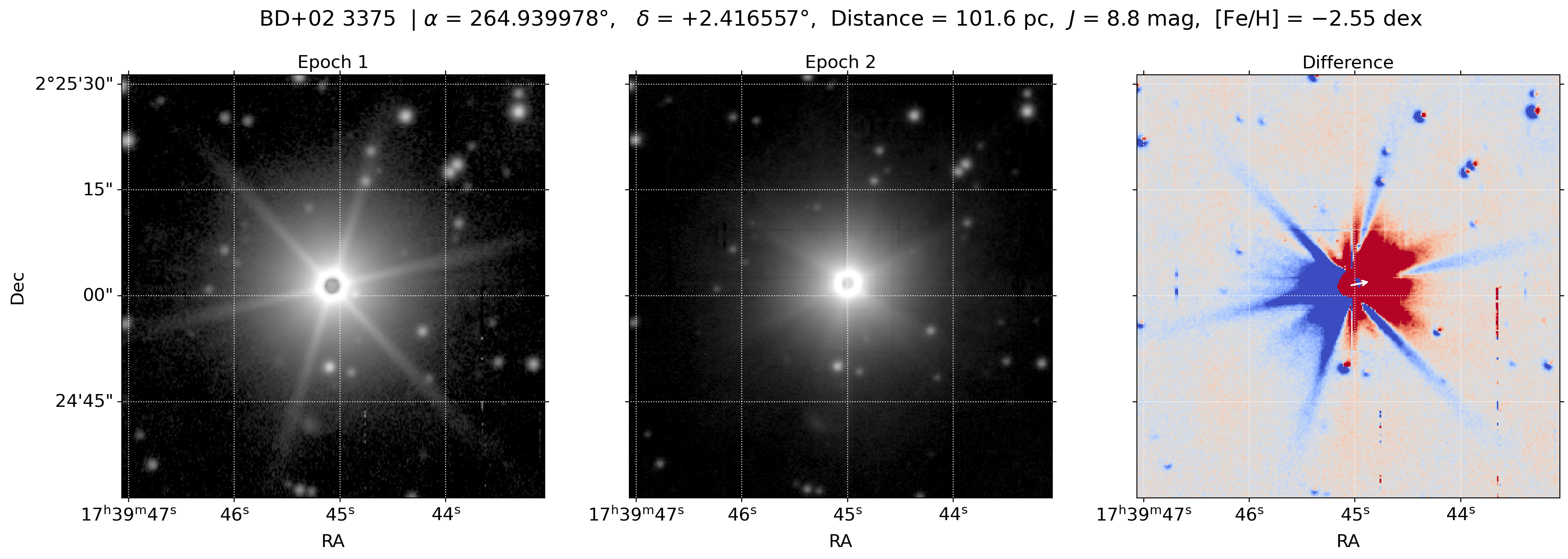}
    \includegraphics[width=0.8\textwidth]{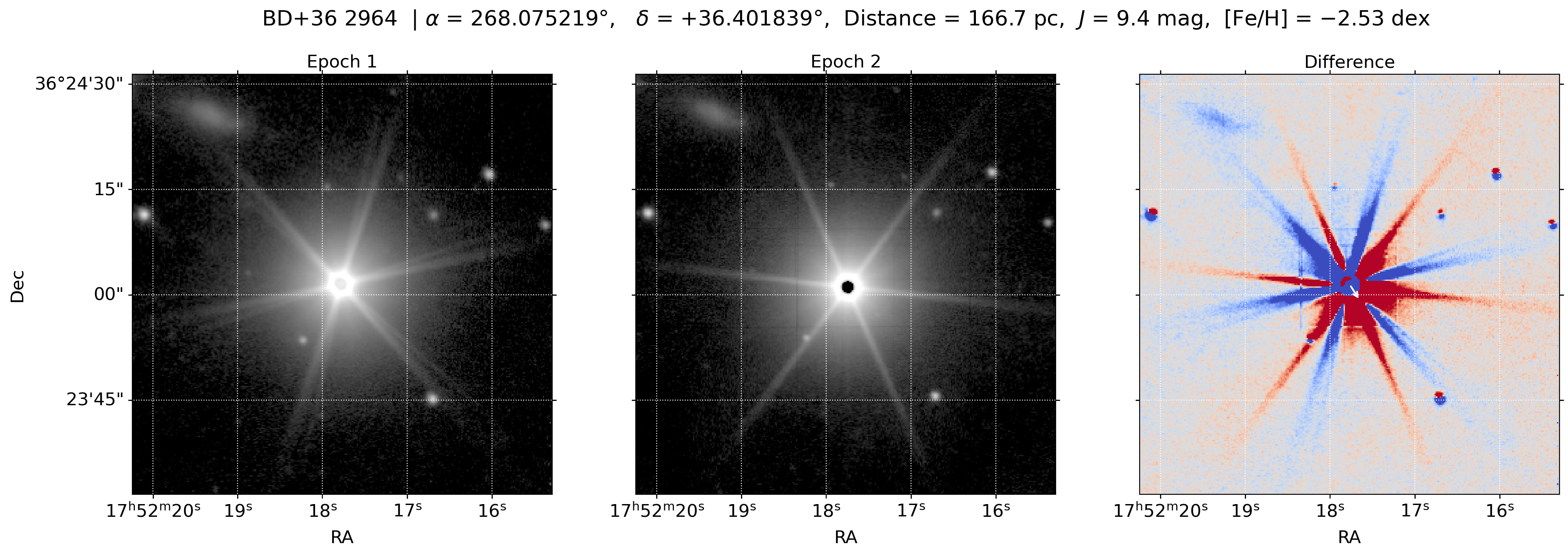}
\end{figure}

\newpage
Figure~\ref{dual} continues:

\begin{figure}[htbp]
\centering
    \includegraphics[width=0.8\textwidth]{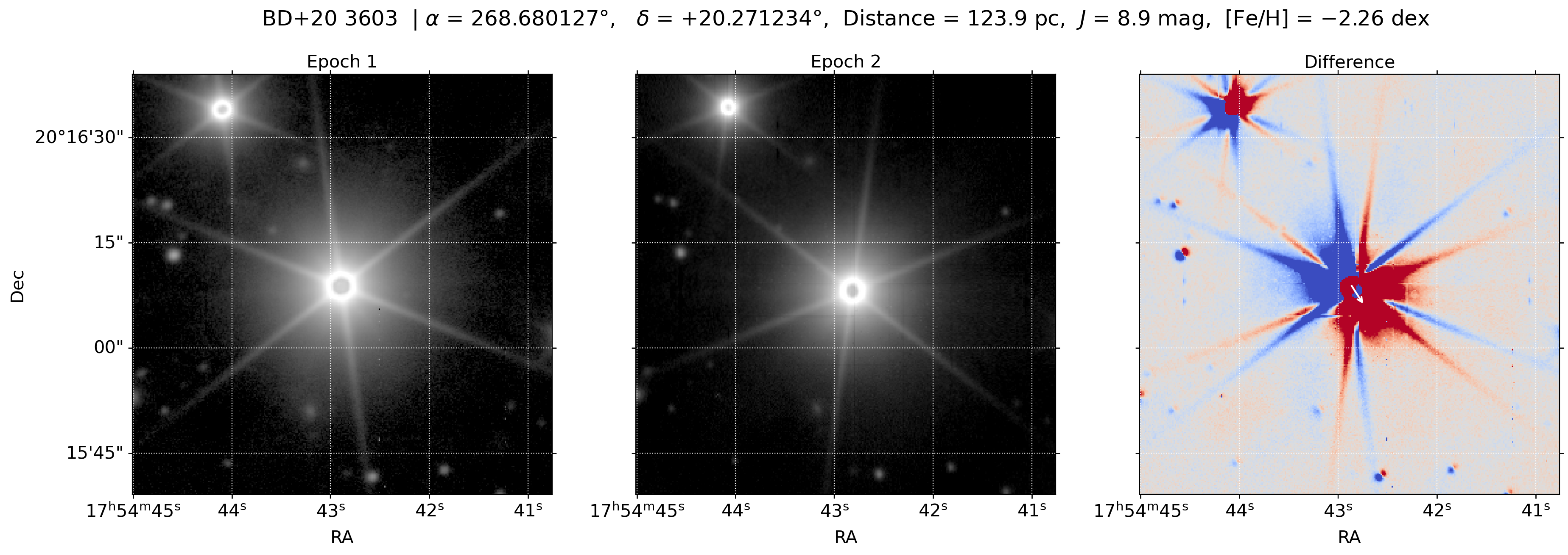}
    \includegraphics[width=0.8\textwidth]{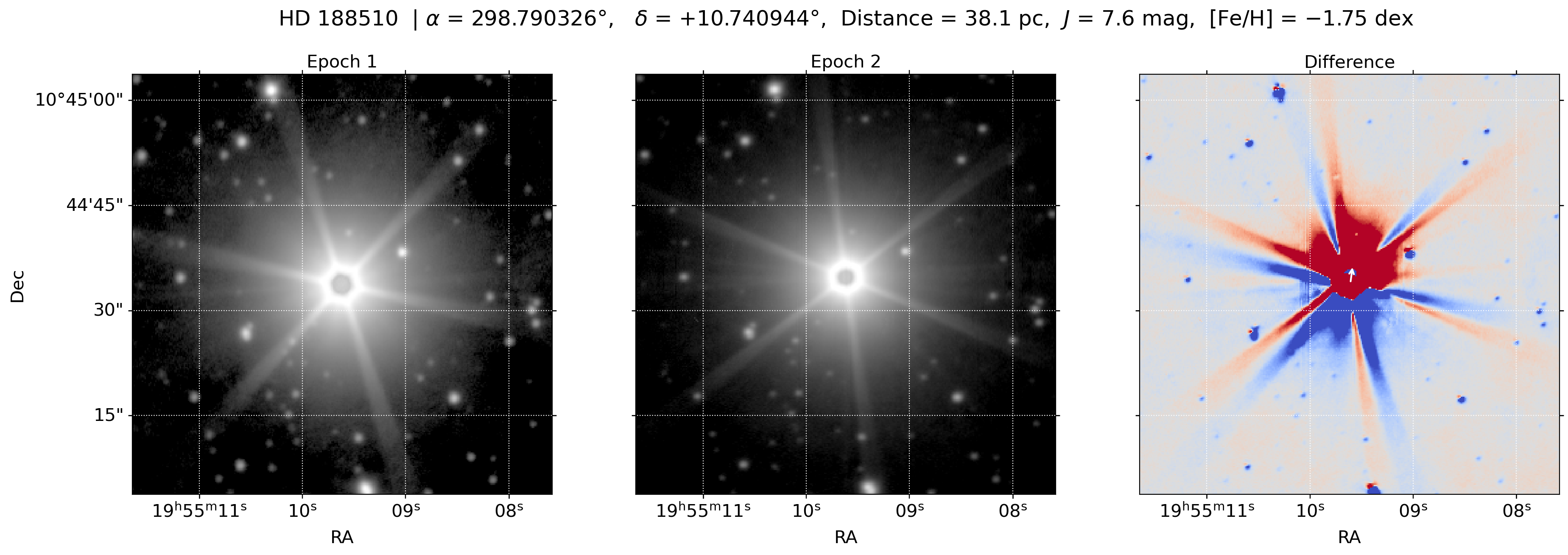}
    \includegraphics[width=0.8\textwidth]{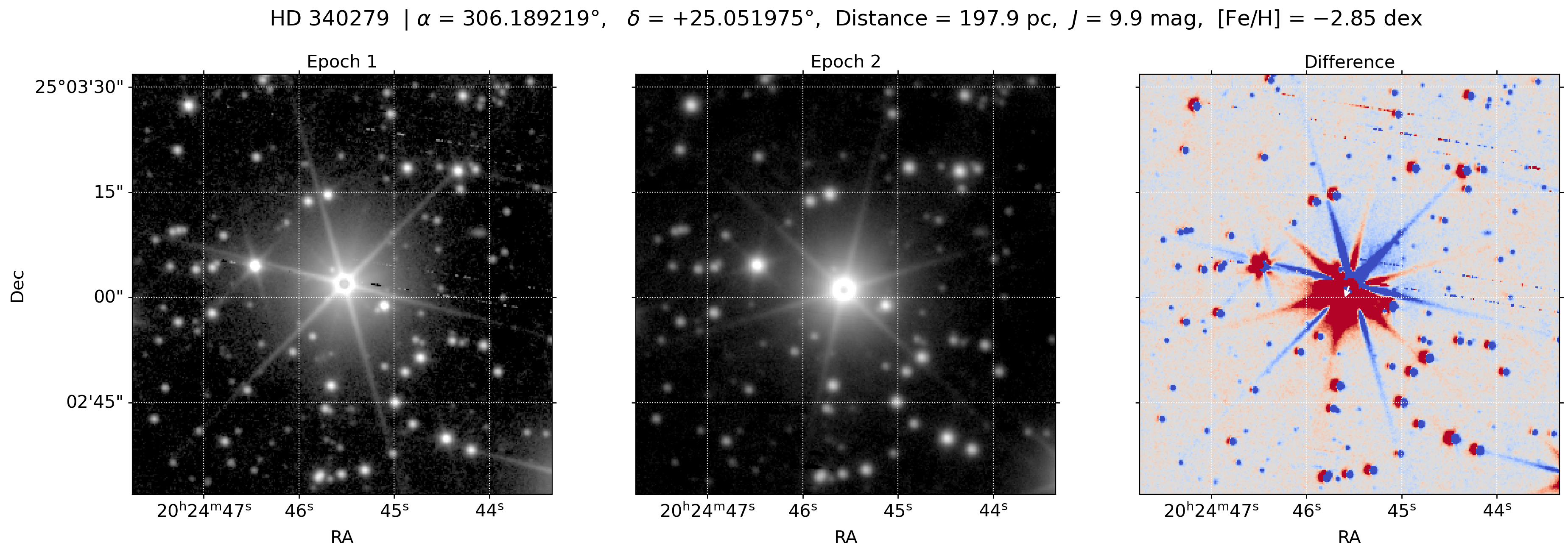}
    \includegraphics[width=0.8\textwidth]{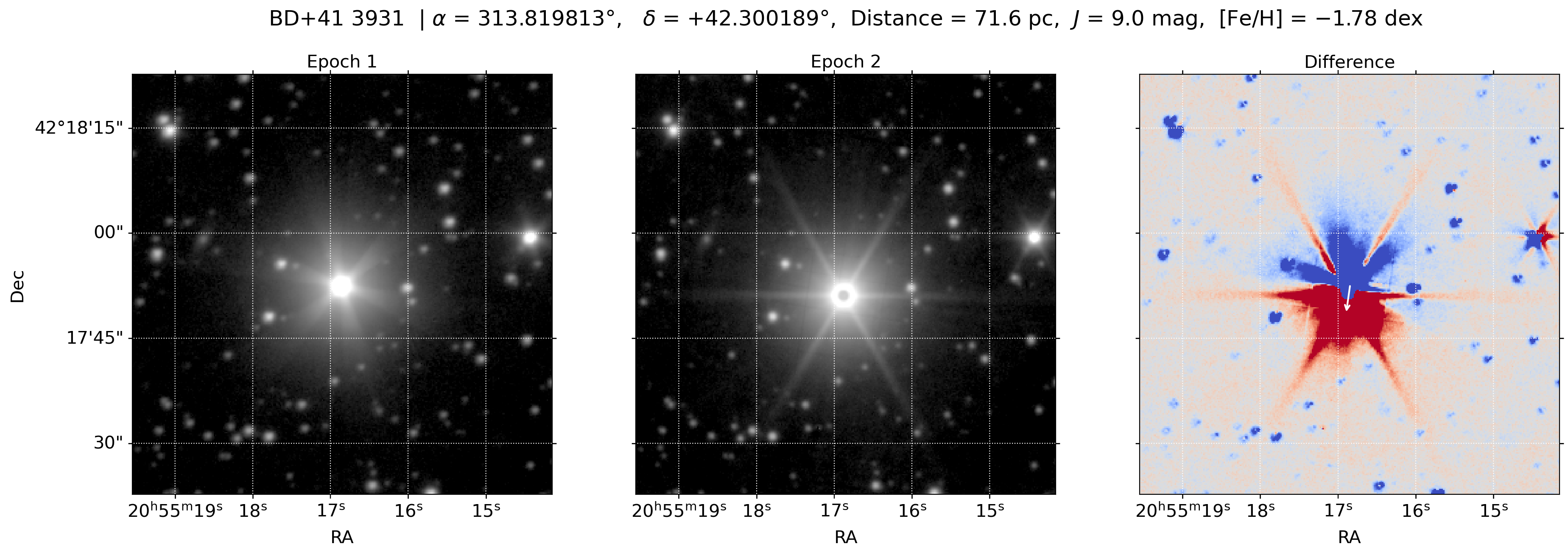}
\end{figure}

\newpage
Figure~\ref{dual} continues:

\begin{figure}[htbp]
\centering
    \includegraphics[width=0.8\textwidth]{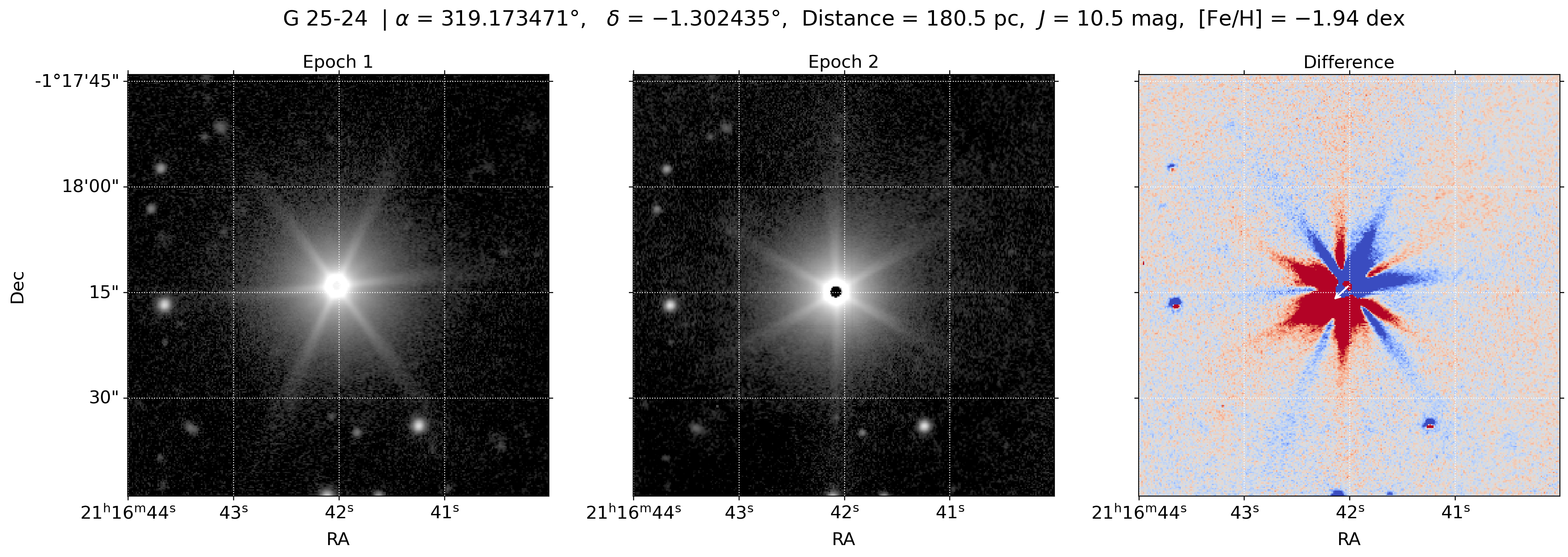}
    \includegraphics[width=0.8\textwidth]{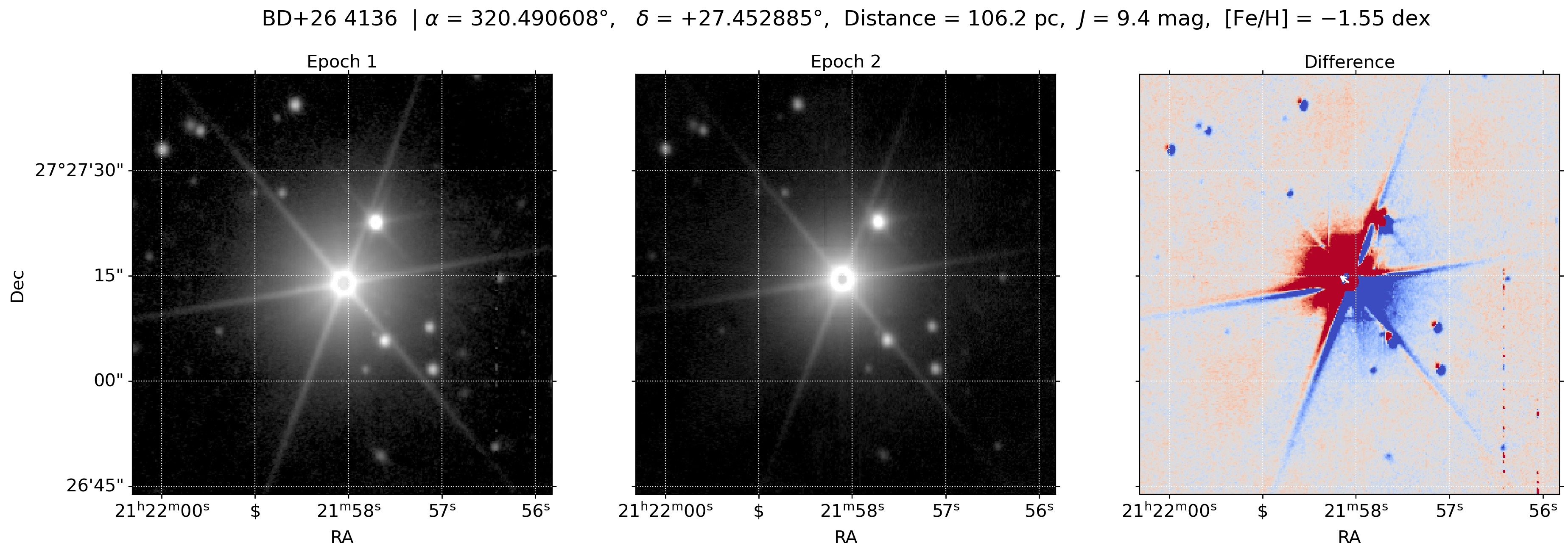}
    \includegraphics[width=0.8\textwidth]{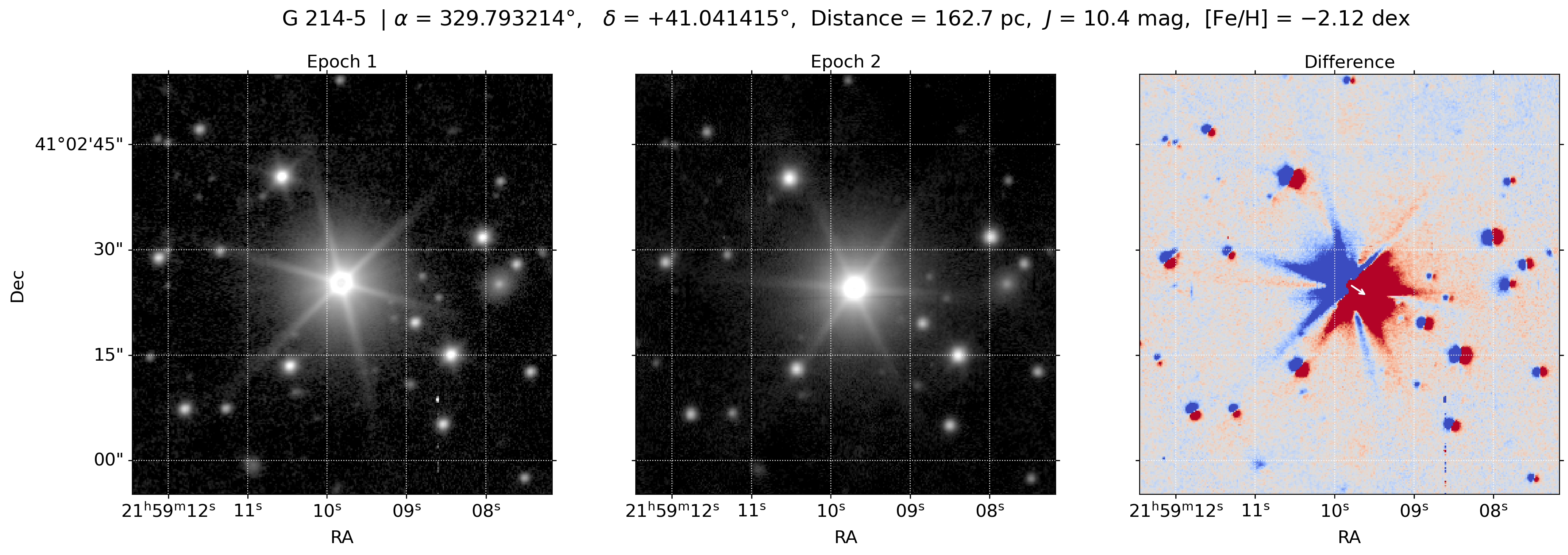}
    \includegraphics[width=0.8\textwidth]{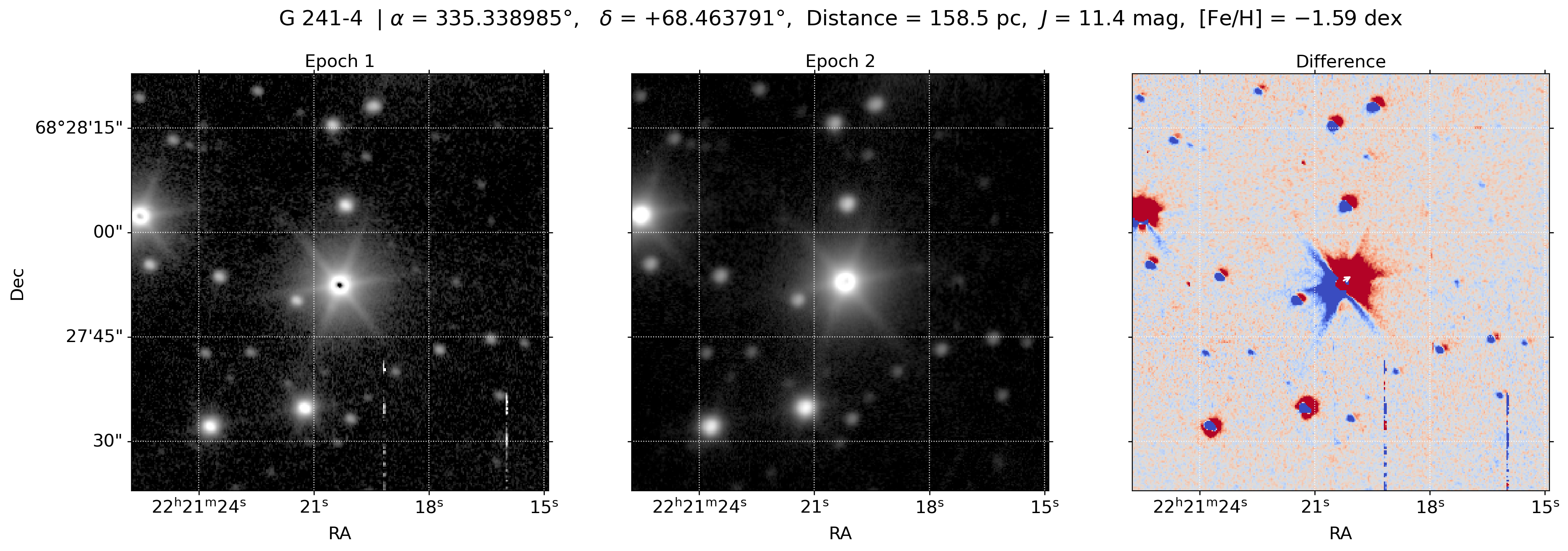}
\end{figure}

\newpage
Figure~\ref{dual} continues:

\begin{figure}[htbp]
\centering
    \includegraphics[width=0.8\textwidth]{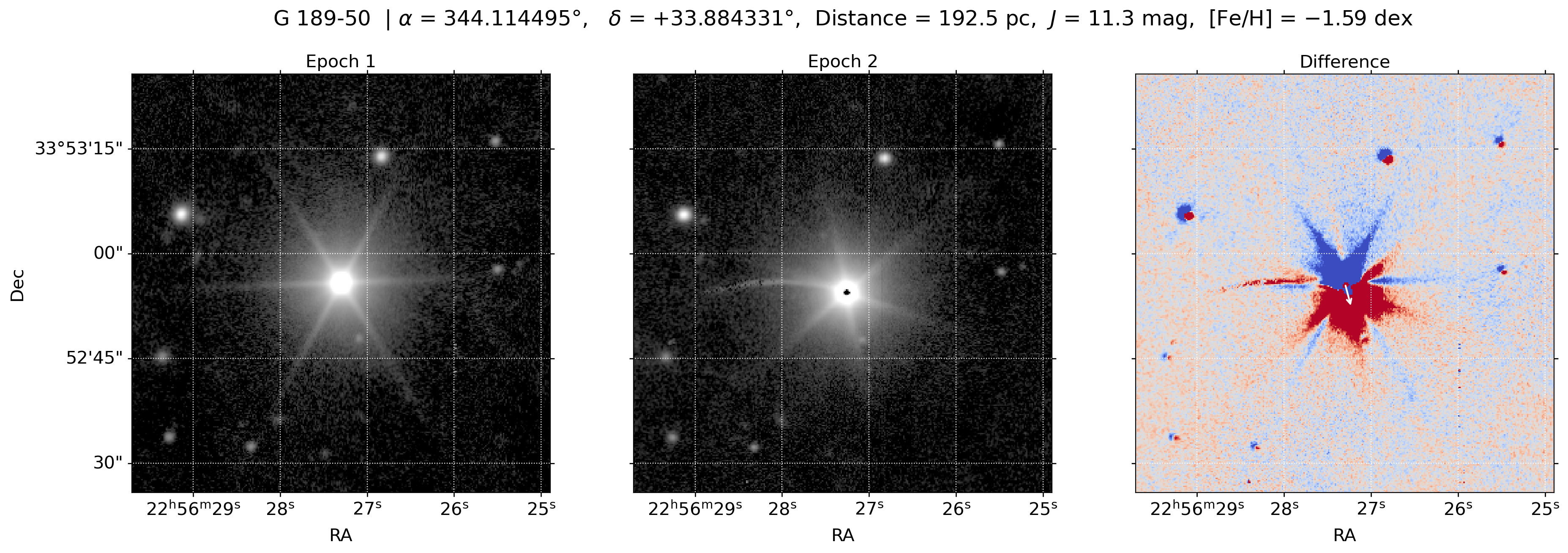}
    \includegraphics[width=0.8\textwidth]{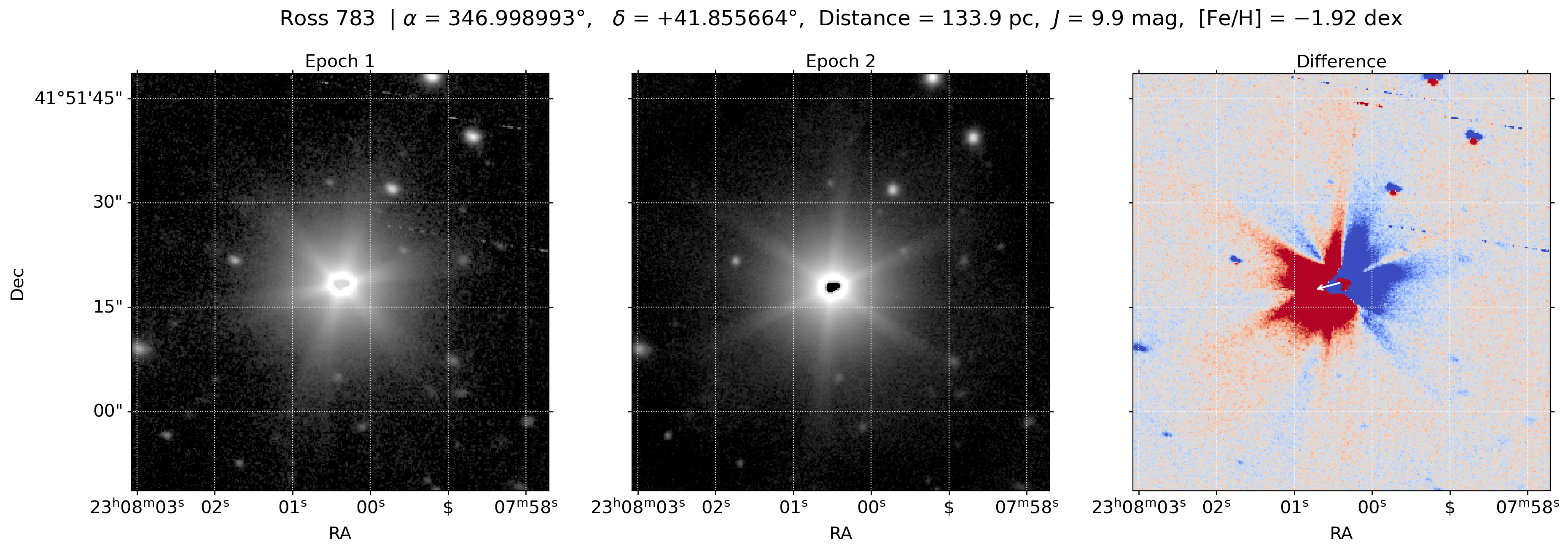}
    \includegraphics[width=0.8\textwidth]{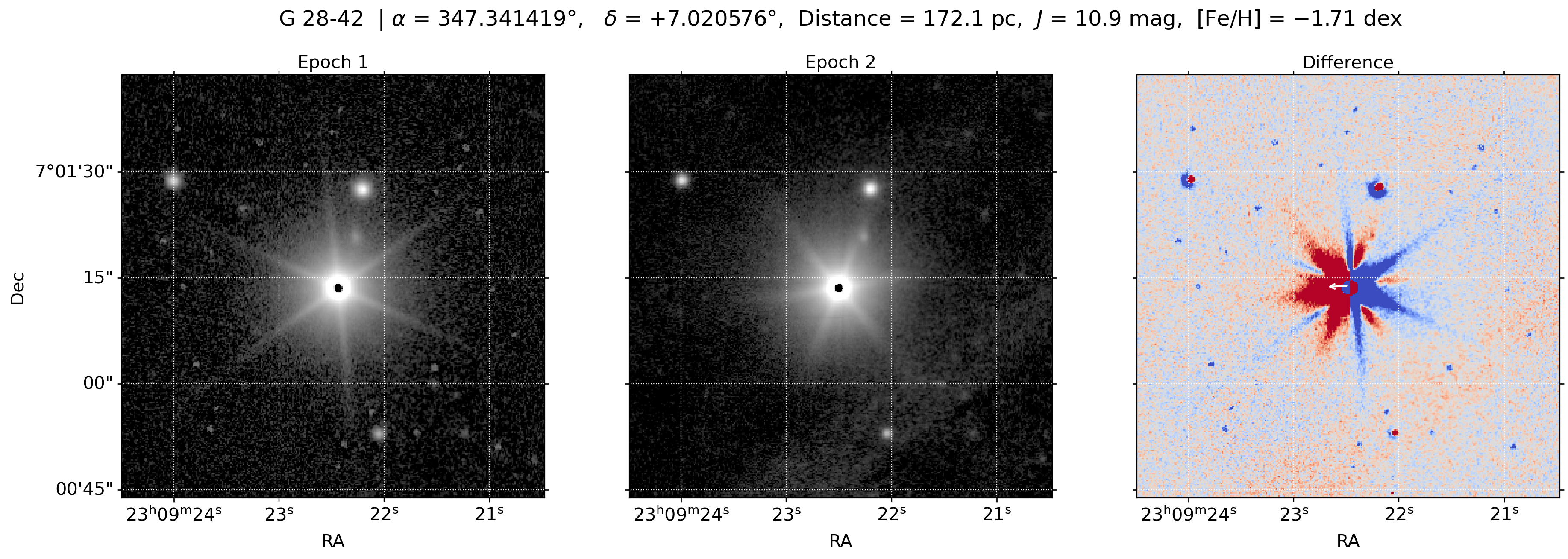}
    \includegraphics[width=0.8\textwidth]{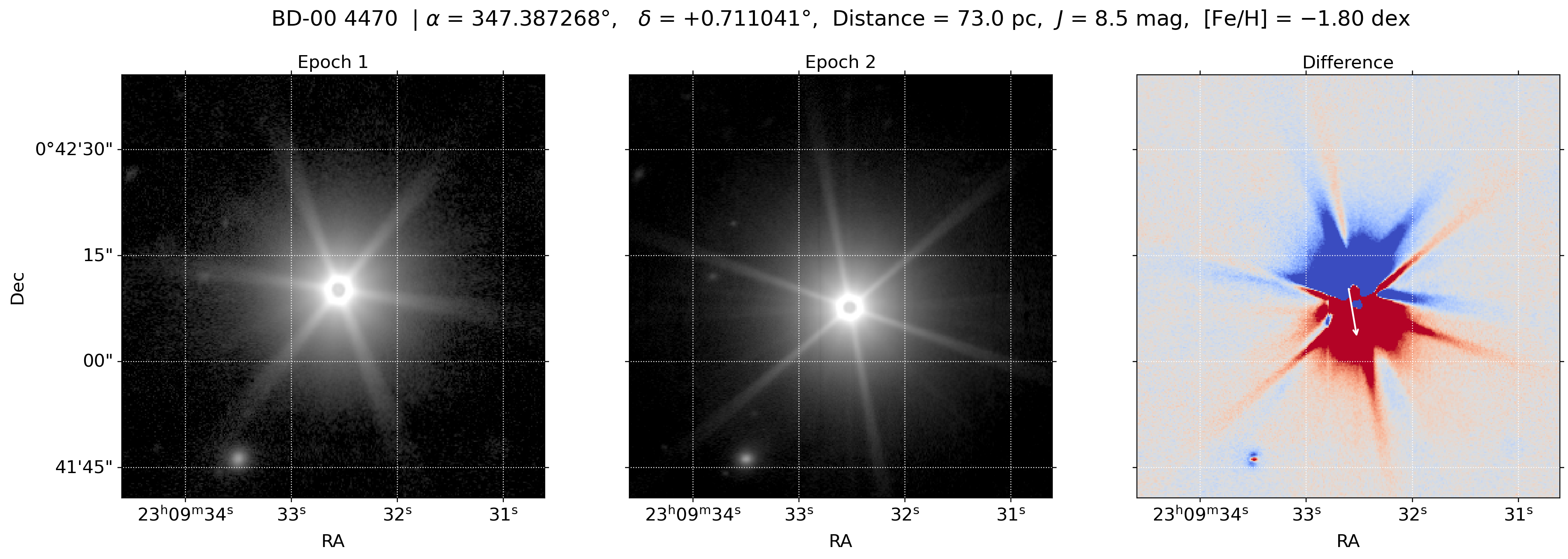}
\end{figure}

\newpage
Figure~\ref{dual} continues:

\begin{figure}[htbp]
\centering
    \includegraphics[width=0.8\textwidth]{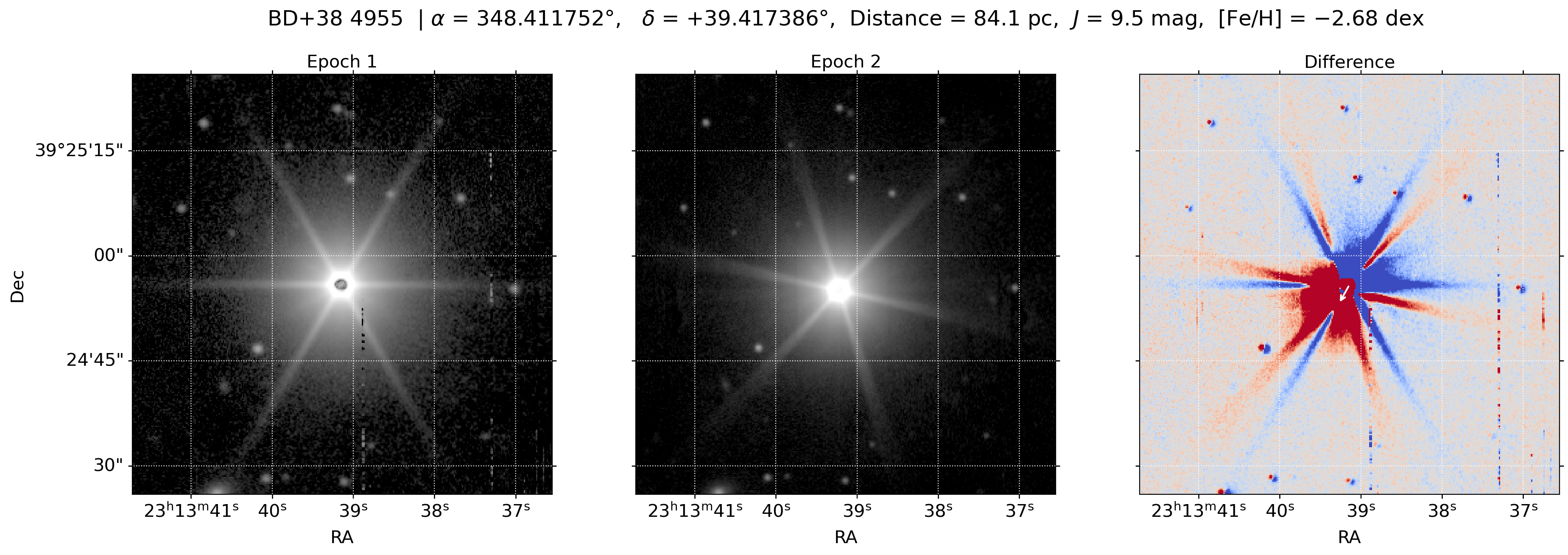}
    \includegraphics[width=0.8\textwidth]{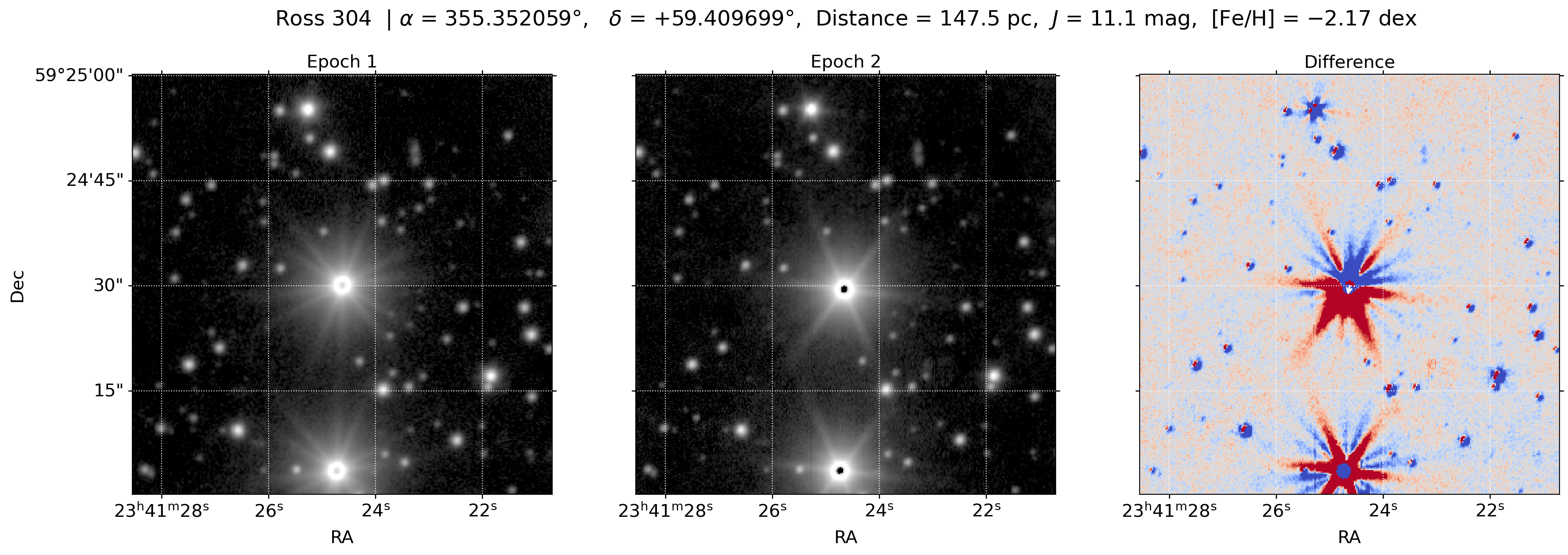}
    \includegraphics[width=0.8\textwidth]{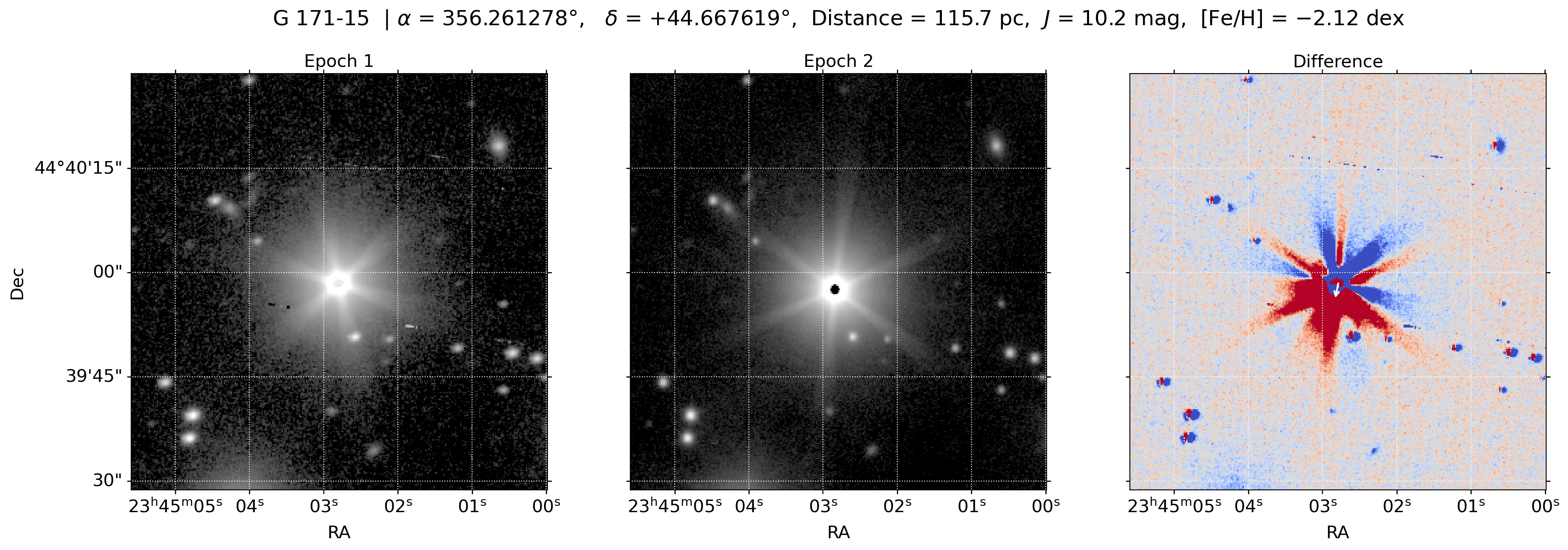}
  \end{figure}
  \clearpage

\end{onecolumn}

\end{appendix}

\endgroup
\end{document}